\documentclass[preprint,authoryear,12pt]{elsarticle}

\usepackage{epsfig}

\usepackage{amssymb}
\usepackage{aas_macros}

\usepackage{color}

\journal{Astronomy and Computing}

\begin{document}

\begin{frontmatter}



\title{{\tt LP-VIcode}: a program to compute a suite of variational chaos
indicators} 


\author[fcag,ialp]{D. D. Carpintero}
\author[fcag,ialp]{N. Maffione}
\author[fcag,ialp]{L. Darriba}

\address[fcag]{Facultad de Ciencias Astron\'omicas y Geof\'\i sicas, Universidad
Nacional de La Plata, Argentina}
\address[ialp]{Instituto de Astrof\'\i sica de La Plata, UNLP-Conicet,
Argentina}

\begin{abstract}

An important point in analysing the dynamics of a given stellar
or planetary system is the reliable identification of the chaotic or regular
behaviour of its orbits. We introduce here the program {\tt LP-VIcode}, a fully
operational code which efficiently computes a suite of ten variational chaos
indicators for dynamical systems in any number of dimensions. The user may
choose to simultaneously compute any number of chaos indicators among the
following: the Lyapunov Exponents, the Mean Exponential Growth factor of Nearby
Orbits, the Slope Estimation of the largest Lyapunov Characteristic Exponent,
the Smaller ALignment Index, the Generalized ALignment Index, the Fast Lyapunov
Indicator, the Othogonal Fast Lyapunov Indicator, the dynamical Spectra of
Stretching Numbers, the Spectral Distance, and the Relative Lyapunov Indicator.
They are combined in an efficient way, allowing the sharing of differential
equations whenever this is possible, and the individual stopping of their
computation when any of them saturates. 

\end{abstract}


\end{frontmatter}

% \linenumbers

\section{Introduction}
\label{intro}

The structure and the dynamics of a self-consistent stellar system are related
to each other; in particular, the orbits of the stars determine the mass
distribution of the system. Thus, in order to model stellar systems, it is of
interest to characterize the orbits supported by the astrophysical potential
giving rise to the model. \citet{S79,S82}, for example, was able to construct
steady state distribution functions of stellar systems starting from a given
potential and a well chosen set of regular orbits supported by the former. Since
these pioneer works, his method has been used by many authors \citep[e.g.][among
many others]{MF96,VVVCD08,DVM11}. This led to the now common claim that regular
orbits constitute a dynamical skeleton for stellar systems. Moreover, those
regular orbits that turn out to be resonant and stable are the most important
ones, because they spawn entire regular families around them; thus, they
constitute the backbone of the system. On the other hand, the existence of
chaotic orbits in realistic models is nowadays beyond doubt
\citep[e.g.,][]{VM98,VKS02,MCW05,DVM11,ZM12}. They are important
to the dynamical evolution of a stellar system because their diffusion through
their allowed phase space may impact the system as a whole
\citep[e.g.][]{MABK95, KS03}; even a new version of the Jeans theorem was
advanced regarding the role of chaotic orbits that the traditional version
ignores completely \citep{K98}. Another interesting point regarding the
relationship between the dynamics of a system and its chaotic orbits is that,
among the latter, those obeying only one isolating integral of motion (dubbed
fully chaotic) and those obeying two such integrals (partially chaotic) occupy
quite different spatial regions \citep{MM04,ZM12}. This suggests different
dynamical roles for each type of chaoticity, though what the roles may be is
still unknown. Although for different reasons, the dynamics of planetary systems
also strongly depends  on the chaoticity of their orbits, in particular, in
terms of the stability of those systems \citep{L90}.

Thus, an important aspect of the dynamical study of a given
stellar or planetary system consists in identifying the chaotic or regular
behaviour of its orbits. Since the early work of \citet{HH64}, the number of
indicators of chaos has steadily grown as time went by. Aside from the
qualitative method of the Poincar\'e surfaces of section \citep[e.g.][]{BT08},
the tools for such analyses are based either on the study of the fundamental
frequencies of the trajectories \citep[e.g.][]{BS82,L90,SN96,CA98,PL98}, or else
on the study of the evolution of deviation vectors, the so-called variational
chaos indicators (CIs hereinafter, see bibliography below). However, among the
available plethora, only a few of them are usually employed by most dynamicists.
Therefore, it would prove useful to have a tool with which one can compute
several CIs in an easy and fast way. This is the main motivation of the present
{\tt LP-VIcode}, which stands for \emph{La Plata Variational Indicators code.}

The alpha version was first introduced in \cite{DMCG12p}. Here, we present a
significantly improved code which constitutes the first stable version of the
program. The main achievement of the code is its speed: neither the orbit nor
any of the sets of variational equations are computed more than once in each
time step, even when they may be requested by more than one CI. The library of
CIs in the present version include the following: the numerically computed
Lyapunov Exponents, known as Lyapunov Indicators, Lyapunov Characteristic
Exponents, Lyapunov Characteristic Numbers, or even Finite-Time Lyapunov
Characteristic Numbers \citep[LIs,][]{BGS76,BGGS80a,BGGS80b}; the Mean
Exponential Growth factor of Nearby Orbits \citep[MEGNO,][]{CS00,CGS03}; the
Slope Estimation of the largest Lyapunov Characteristic Exponent
\citep[SElLCE,][]{CGS03}; the Smaller ALignment Index \citep[SALI,][]{S01}; the 
Generalized ALignment Index \citep[GALI,][]{SBA07}; the Fast Lyapunov Indicator
\citep[FLI,][]{FGL97,LF01}; the Orthogonal Fast Lyapunov Indicator 
\citep[OFLI,][]{FLFF02}; the Spectral Distance \citep[SD,][]{VCE99}; the
dynamical Spectra of Stretching Numbers \citep[SSNs,][]{VC94,CV96}; and the
Relative Lyapunov Indicator \citep[RLI,][]{SEE00,SESF04}.\footnote{A minimal
package of CIs for analysing a general Hamiltonian system is studied in
\citet{MDCG11,DMCG12,MDCG13}.} The potential of the dynamical system, which must
be supplied by the user together with the corresponding accelerations and
variational equations, may have any number of dimensions (or degrees of
freedom). The user may then choose to compute simultaneoulsy any subset of the
abovementioned CIs.

As already stated, the main goal of the program is to compute, for one or more
orbits in a given potential, the set of chosen CIs at the same time and in an
efficient way. This means that whenever two or more CIs need the same
differential equations, the last would be integrated only once. Thus, at each
run of the program the set of equations adaptes to the set of CIs the user has
chosen to compute.

\section{Structure of the code}

The program is written in standard Fortran77\footnote{There is also a version
written in Fortran90 for parallel computing, though in a developing stage.},
except for the common nonstandard extensions {\tt DO-ENDDO}, {\tt INCLUDE}, {\tt
DOWHILE}, lowercase characters, inline comments, names longer than 6 characters,
and variable names containing the nonstandard character "\_". All real variables
are {\tt DOUBLE PRECISION}.

The program reads initial conditions (hereinafter, i.c.) for one or more orbits,
and integrates them, computing at the same time the set of CIs chosen by the
user. The integrator is a standard Bulirsch-Stoer routine. The
fixed time step asked by the user is internally split whenever either the
absolute or relative error of the step is greater than the allowed tolerance,
set to $10^{-13}$ for our experiments except for the MEGNO, for which the
tolerance was set to $10^{-12}$. This difference turns out to be not important
at all for the MEGNO, but avoids the slowing down of the code when this CI is
combined with any other.

The code is organized as a main program which controls the flux of the
computation, plus a set of routines to accomplish the different tasks. These
routines are grouped into four categories: those which take care of the input,
initialization and output, those which compute the CIs themselves, those which
are mathematical tools, and those which deal with the dynamics of the system.
This last category includes the routines the user must supply, i.e., those that
correspond to the potential being studied.

\subsection{The main program}

The main program begins with a call to the input routine, after which the main
loop sweeps the different orbits. Inside this loop, the differential equations
are adjusted to the set of CIs to be computed. Then, any needed deviation vector
(DV) is generated, and the equations of motion as well as the variational
equations are integrated. At each step, only one call to the integration routine
is done, irrespective of the CIs being computed. After that, the chosen CIs are
computed. If any of the CIs saturates during the integration of an orbit, the
last value of the former is output, and the corresponding equations are deleted
if none of the other CIs is using them. This continues until the integration
loop is finished, in which case a new set of i.c. is read, the differential
equations are reset for the new orbit, and the whole process is repeated.

\subsection{Input, initialization, and output}

The user provides the input parameters (time step, length of the integration,
choice of the CIs, frequency of the output, where to find the i.c., etc.)
through a file.  Although not strictly an input from the user, he/she should
also provide the routines with which the potential, the accelerations and the
variational equations should be computed. The syntax of these routines, as well
as the structure of the input file and of the possible outputs, are described in
detail in the User's Guide provided with the program.

The initialization is a three-step process. First, the DVs are generated.
Second, the initial phase space values and the initial deviation vectors
(hereinafter IDVs) are stored in a matrix which therefore holds all the
dependent variables of the problem. Finally, according to the chosen CIs, the
bookkeeping of equations and DVs to be used is done (more on this below).

\subsection{Indicators}\label{Indicators}

We briefly introduce here the corresponding definitions and algorithms of the
indicators used in the code. Consider the Hamiltonian $H(\mathbf{w})$, with
$\mathbf{w} = (\mathbf{p}, \mathbf{q})$ a phase-space  vector, {\bf q} the
position of the orbit and {\bf p} its momentum. Introducing the function
$\mathbf{F}$:
\begin{equation}
\mathbf{F}(\mathbf{w})=(\partial H/\partial\mathbf{q},\ -\partial
H/\partial\mathbf{q}),
\end{equation}
the equations of motion can be written as
\begin{equation}
\dot\mathbf{w}=\mathbf{F}(\mathbf{w}).
\label{eqmot}
\end{equation}

Solving this system of first order ordinary differential equations
with i.c. $\mathbf{w}_0$, we obtain the solution 
\begin{equation}
\mathbf{w}(t) = \mathbf{\Phi}^t\mathbf{w}_0,
\label{solution}
\end{equation}
where $\mathbf{\Phi}^t$ is the operator evolution. Now, taking the first
variation of Eqs. (\ref{solution}), we obtain the so-called variational
equations
\begin{equation}
\delta\mathbf{w}(t)=\mathrm d_{\mathbf{w}}\mathbf{\Phi}^t\delta\mathbf{w}_0,
\label{varec}
\end{equation}
where $\delta\mathbf{w}$ is a DV and d$_{\mathbf{w}}$ stands for the operator
derivative with respect to the components of {\bf w}.

\subsubsection{The LI, the MEGNO and the SElLCE}
\label{LIMEGNOSElLCE}

We can gain fundamental information about the Hamiltonian flow in the
neighborhood of any orbit through the so-called LIs
\citep{BGS76,BGGS80a,BGGS80b}: 
\begin{equation}
\mathrm{LI}_j(t) =
\frac{1}{t}\ln\frac{\|\delta\mathbf{w}_j(t)\|}
{\|\delta\mathbf{w}_{j0}\|},
\end{equation}
where $\|\cdot\|$ is some norm (usually the Cartesian one), and the subindex $j$
indicates that the calculation is to be performed using the $j$th deviation
vector. For a system of $n$ dimensions, there will be $2n$ LIs, $n$ of which
will be positive and the other $n$ will be their respective opposites. Since,
before the actual computation, the deviation vectors should be orthonormalized
\citep{BGGS80a,BGGS80b}, we do this in the code by means of a Modified
Gram-Schmidt procedure. It is worth noticing that, having tried both the
discrete and continuous Eckmann and Ruelle algorithms of orthogonalisation
\citep{ER85}, we have decided to discard them because, in the first case
(discrete), the integration time within which the results were reliable turned
out to be limited, whereas in the second case (continuous) the computing time
was about 250 per cent that of the Gram-Schmidt procedure.

Among the set of LIs, the largest one is the most important, since it suffices
to determine whether an orbit is regular or chaotic: if it tends to a positive
value, then the orbit is chaotic, regardless of the behaviour of the rest of the
LIs. On the other hand, if the largest LI tends to zero, the rest of LIs will
also tend to zero, and the orbit in this case is regular. Hereinafter, we will
refer to the largest LI just as LI.

The LI is the
truncated (in time) value of the largest Lyapunov Characteristic Exponent
$\sigma_1$ that can be defined in an integral form as:
\begin{equation}
\sigma_1=\lim_{t\to\infty}{1\over t}\int_0^t{\|\dot\delta\mathbf{w}(t')\|
\over\|\delta\mathbf{w}(t')\|}\mathrm{d}t',
\end{equation}
where $\dot\delta\mathbf{w}$ is a shorthand for
$\mathrm{d}(\delta\mathbf{w})/\mathrm{d}t'$. Both $\sigma_1$ and the LI tend to
a positive value for chaotic orbits, and  tend to 0 for regular orbits. 

Now we introduce the MEGNO $Y$ \citep{CS00,CGS03} through the expression:
\begin{equation}
Y(t)={2\over
t}\int_0^t{\|\dot\delta\mathbf{w}(t')\|\over\|\delta\mathbf{w}(t')\|}t'
\mathrm{d}t'.
\end{equation}

The actual expression of the indicator used in the code is its time average:
\begin{equation}
\overline{Y}(t)={1\over t}\int_0^t Y(t')\,\mathrm{d}t'.
\end{equation}
The asymptotic behavior of $\overline{Y}$ can be described as
$\overline{Y}(t)\to a\cdot t+b$, where $a=\sigma_1/2$ and $b=0$ for irregular,
stochastic motion, while $a=0$ and $b=2$ for quasi-periodic motion. It is also
possible to estimate the LI of the orbit from $a\cdot t+b$  by applying a linear
least-squares fitting on $\overline{Y}(t)$. This estimation is the SElLCE
indicator \citep{CGS03}, in the computation of which only the last 80 per cent
of the orbit is used, in order to avoid any initial transient.

\subsubsection{The SALI and the GALI}
\label{SALIGALI}

Let $\delta\mathbf{w}_1$ and $\delta\mathbf{w}_2$ be two DVs belonging to the
same orbit, linearly independent at $t=0$, and let $\delta\hat\mathbf{w}_i(t) =
\delta\mathbf{w}_i(t) / \|\delta\mathbf{w}_i(t)\|$, $i=1,2$ be the corresponding
unit DVs. The parallel and antiparallel alignment indices are then defined as
$d_- = \|\delta\hat\mathbf{w}_1 - \delta\hat\mathbf{w}_2\|$ and $d_+ = 
\|\delta\hat\mathbf{w}_1 + \delta\hat\mathbf{w}_2\|$, respectively \citep{S01}.
Since, in a chaotic motion, two linearly independent DVs are expected to align
with the same direction, we will have $d_-\rightarrow 0$ and $d_+\rightarrow 2$
or $d_-\rightarrow 2$ and $d_+\rightarrow 0$. On the other hand, if the motion
is regular, $d_-$ and $d_+$ will oscillate within the interval $(0,2)$. The SALI
\citep{S01} is defined as the smaller of these two indices:
\begin{equation}
\mathrm{SALI}(t)=\mathrm{min}(d_+,d_-),
\end{equation}
so that SALI $\to 0$ if the orbit is chaotic, whereas SALI $\not\to 0$ if the
orbit is regular.

\citet{SBA07} generalize the SALI introducing an alternative way to compute it.
They evaluate the wedge product $\|\delta\hat\mathbf{w}_1 \wedge 
\delta\hat\mathbf{w}_2\|\equiv (d_+\cdot d_-)/2$, that is, the area of the
parallelogram formed by the two DVs, which has the same behaviour as the SALI
for regular and chaotic motions. Taking more independent IDVs, the wedge product
can be generalized up to $k$ factors, $2< k\le 2n$, with $n$ the number of
degrees of freedom, representing the volume of the parallelepiped formed by the
$k$ DVs. The GALI$_k$ is defined as the volume of this $k$-parallelepiped:
\begin{equation}
\mathrm{GALI}_k(t) = \|\delta\hat\mathbf{w}_1(t)\wedge\delta\hat\mathbf{w}_2(t)
\wedge\ldots\wedge\delta\hat\mathbf{w}_k(t)\|.
\label{gali1}
\end{equation}

The implementation of Eq. (\ref{gali1}) in the code is made through a singular
value decomposition of the matrix of DVs \citep{SBA08}. As in the case of the
SALI, the GALI$_k$ for chaotic orbits tend to zero, although giving more
information about the dynamics of the orbit than the SALI. On the other hand,
if GALI$_k$ tends to a non-zero value as $t$ increases, the motion is regular.

\subsubsection{The FLI and the OFLI}
\label{FLIOFLI}

Given $n$ linearly independent DVs in a $2n$ dimensional phase space, the FLI
\citep{FGL97,LF01} at time $t$ is defined as the greatest  of the norms they had
between $t=0$ and the current $t$:
\begin{equation}
\mathrm{FLI}(t)=\mathrm{sup}_t\left[\|\delta\mathbf{w}_1(t)\|,
\|\delta\mathbf{w}_2(t)\|,\ldots,\|\delta\mathbf{w}_n(t)\|\right].
\end{equation} 
It turns out that the FLI grows exponentially for chaotic motion and linearly
for regular motion.

The OFLI \citep{FLFF02}, on the other hand, is computed like the FLI, but only
the component orthogonal to the flow of each DV is taken into account. This
modification makes the OFLI a CI that can easily distinguish periodicity of a
regular motion: in this case, it oscillates around a constant value, while for
chaotic and quasiperiodic motion it has the same behavior as the FLI.

\subsubsection{The SSNs and the SD}
\label{SSNSD}

The stretching number $s_i$ \citep{VC94,CV96} is defined as:
\begin{equation}
s_i=\frac{1}{\Delta t}\ln\frac{\|\delta\hat\mathbf{w}(t_i)\|}
{\|\delta\hat\mathbf{w}(t_{i-1})\|},
\end{equation}
where $\Delta t$ is the step of integration, and $t_i=i \Delta t$,
$i\in\mathbb{N}$. The spectrum of the $s_i$, called SSN, is defined as the
probability density of their values. If the $s_i$ are binned into  blocks of
width $\Delta s$, then the SSN can be computed as:
\begin{equation}
\mathrm{SSN}_j(t)={1\over Z}\frac{\Delta Z_j}{\Delta s},\quad
j=1,\dots,N,
\end{equation}
where $N$ is the number of blocks of the histogram, $Z$ is the total number of
$s_i$ and  $\Delta Z_j$ is the number of $s_i$ in the $j$-th interval. However,
when the sample is very large, the analysis of the orbits by means of the SSN is
no longer reliable. Thus, \citet{VCE99} introduce the SD:
\begin{equation}
\mathrm{SD}^2(t)=\sum_{j=1}^N [\mathrm{SSN}_{j,1} - 
\mathrm{SSN}_{j,2}]^2\cdot \Delta s,
\end{equation}
where SSN$_{j,i}$ is the  SSN$_j$ corresponding to a DV
$\delta\hat\mathbf{w}_i$. If the orbit is chaotic, the SD decreases towards
zero. If, instead, the orbit is regular, the SD tends to a constant non-zero
value.

\subsubsection{The RLI}\label{RLI}

Let $\mathbf{w}(t)$ be an orbit with i.c. $\mathbf{w}_0$ (the \emph{base
orbit}), and $\mathbf{w}(t)+ \Delta \mathbf{w}(t)$ be another orbit with i.c.
$\mathbf{w}_0+ \Delta \mathbf{w}_0$ (the \emph{shadow orbit}), where $\Delta
\mathbf{w}_0$ is small. From the LIs of these orbits LI$_0(t)$ and LI$_1(t)$,
respectively, the RLI \citep{SEE00,SESF04} is defined as:
\begin{equation}
\mathrm{RLI}(t)=\|\mathrm{LI}_1(t)-\mathrm{LI}_0(t)\|.
\end{equation}

Since the LI usually has fast fluctuations, it is better to smooth those out by
averaging with respect to time, so the RLI is usually redefined and computed as
\begin{equation}
\overline\mathrm{RLI}(t)=\frac{1}{t}
\sum_{i=1}^{t/\Delta t}\mathrm{RLI}(t_i),
\end{equation}
where $\Delta t$ is the step of integration and $t_i=i \Delta t$,
$i\in\mathbb{N}$. The RLI values for chaotic motion exceed by several orders of
magnitude those associated with regular motion.

\subsection{Saturation}

After the requested CIs are computed, the code takes care of any possible
saturation: since there are CIs that tend to infinity when chaos is present,
they are not further integrated after reaching a given threshold in order to
avoid overflows ($10^{16}$ in the case of the FLI and the OFLI, and 30 in the
case of the MEGNO). Similarly, there are CIs (SALI and GALIs) that tend to 0
when chaos is present; for these CIs, we stop their computation when a value of
$10^{-16}$ is reached, i.e., we take the machine precision as a numerical zero.
In all cases, the differential equations are adapted to the new set of CIs by
removing those which correspond to the saturated ones.

\subsection{Mathematical and dynamical routines}

The purely mathematical routines are standard ones; some of them are adaptations
of those published in \citet{PTVF92}. The dynamical routines comprise two sets:
those already written down, and those that must be provided by the user. In the
first group, besides the computation of the energy, there is a routine which
sets all the equations of motion and variational equations that are the same
irrespective of the potential used. These equations are the following. Let ${\bf
x}$ be the position vector in an $n$-dimensional space. The equations of motion
corresponding to some potential $\Phi$ on this space are:
\begin{equation}
\ddot{\bf x} = -\nabla\Phi({\bf x}),
\end{equation}
which, for the sake of their numerical integration, are usually broken into
the first-order differential equations
\begin{eqnarray}
\dot{\bf x} &=& {\bf v},             \label{vel}\\
\dot{\bf v} &=& -\nabla\Phi({\bf x}), \label{acc}
\end{eqnarray}
where {\bf v} is the velocity vector. As can be seen, Eqs. (\ref{vel}) contain
no references to the potential, so we have already coded them into our program.
The same goes for the variational equations. Let ${\bf w}= ({\bf x},{\bf v})$ be
a phase-space vector, and let {\bf F} be such that Eqs. (\ref{vel}) and
(\ref{acc}) are written as  in Eq. (\ref{eqmot}). Then, the variational
equations are
\begin{equation}
\frac{\rm d (\delta{\bf w})}{\rm d t} = \left.\frac{\partial {\bf F}}
{\partial {\bf w}}\right|_{\bf w} \delta{\bf w},
\label{vareq}
\end{equation}
which are the same as Eq. (\ref{varec}), but now written in terms of the
function {\bf F}. Again, the first $n$ equations, namely
\begin{equation}
\dot\delta\mathbf{x}=\delta\mathbf{v}
\end{equation}
are independent of the potential, so we have wired them into the code.

The routines the user must supply should only contain, then, the $n$ equations
of motion (\ref{acc}) and the $n$ variational equations (\ref{vareq}) which do
depend on the potential, plus a routine to compute the value of the potential
itself. The User's Guide contains a complete example which can be used as a
template.

\section{Management of the differential equations}

We took advantage of the passing-by-address paradigm of the Fortran77 language
to manage the equations to be integrated according to the CIs being computed,
i.e., those requested minus those that have saturated.

To this end, we first split the CIs into three groups, according to the type of
DVs they use. These groups are:
\begin{enumerate}[(I)]
\item Normalized and orthogonalized DVs (LI).
\item Normalized DVs (SALI, GALI, SD, SSN, RLI; the RLI needs also a second
(shadow) orbit with its own normalized DV.)
\item Unnormalized DVs (MEGNO, SElLCE, FLI, OFLI).
\end{enumerate}

\begin{figure}
\epsfxsize=\hsize \epsfbox{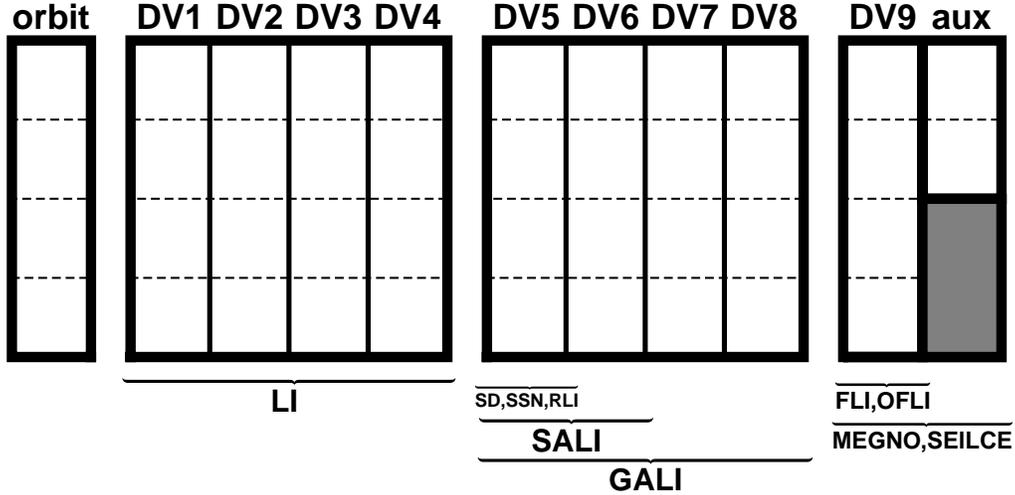}
\label{xva}
\caption{Matrix of dependent variables, for a 2D potential (i.e., 4D phase
space). Rows are coordinates of the phase space, columns are the various sets of
dependent variables, that is, orbit and DVs. The columns are separated into the
three groups of DVs defined in the text. Here, the matrix corresponds to all the
CIs switched on.}
\end{figure}

Then, we set up a $N\times (M+1)$ matrix (Fig. \ref{xva}), where $N$ is the
dimension of the phase space (which depends on the potential being studied), and
$M$ is the total number of DVs to be integrated, which is computed from the
requested CIs as follows. If the group (I) is to be computed, it requires $N$
DVs. In group (II), one DV is needed if only the SD and/or the SSNs and/or the
RLI are sought (the second orbit of the RLI is not counted in; an additional DV
is integrated separately for the shadow orbit), two DVs if the SALI is present,
or $N$ DVs if the GALIs enter the computation. If there are a combination of
these CIs to be computed, only the greater number of DVs are set up; the CIs
requiring a number of DVs less than this maximum will share the DVs with the
others. Finally, group (III) asks for only one DV, irrespective of the number of
indicators present. But, if the MEGNO or the SElLCE are requested, an additional
column is set up to hold the two additional dependent variables of the
differential equations that are needed to compute those CIs (marked as {\tt aux}
in Fig. \ref{xva}). Note that these last two dependent variables remain two in
number irrespective of $N$. The remaining (first) column of the matrix holds the
coordinates of the orbit itself.

If, for example, the potential is 2D (as in Fig. \ref{xva}), and the GALI is not
computed, the program eliminates the DVs that belongs exclusively to this CI
(DV7 and DV8), leaving the other two of group (II) for the computation of the
other CIs. When a CI which is being computed saturates, the program also
eliminates its DVs, though only if they are not being used by another CI. In
this way, the computing time is fully optimized.

\section{Results}

We chose two astronomical potentials, the 2D H\'enon-Heiles \citep{HH64} and a
3D triaxial NFW profile \citep{VWHS08}, in order to show the performance of the
code, as well as some additional characteristics. In all cases we used an Intel
Core i5 with four cores, CPU at 2.67 GHz, 3 GB of RAM, an OS of 32 bits, and the
{\tt gfortran} compiler of {\tt gcc} version 4.4.4, without any optimizations.

\subsection{The H\'enon-Heiles potential}

For the H\'enon-Heiles potential, we took three sets of i.c. For each of these
sets, we recorded the elapsed times in computing all the indicators both
together and individually, in order to assess the saved time. 

The first set, dubbed H1, consisted of four orbits lying on the $x=p_y=0$ line
\citep{CGS03}. Table \ref{tcih1} shows the i.c., which correspond to a stable
periodic orbit {\it sp}, a quasi-periodic orbit {\it qp}, a quasi periodic orbit
near an unstable periodic orbit {\it up}, and a chaotic orbit {\it c1},
all lying on the $E=0.118$ energy surface.

\begin{table}
 \caption{Initial conditions for the experiment H1.}
 \label{tcih1}
 \begin{center}
 \begin{tabular}{lllll}
  \hline
  Name & $x$ & $y$ & $p_x$ & $p_y$ \\
  \hline
  \emph{sp} &0 & 0.295456 & 0.407308431 & 0 \\
  \emph{qp} &0 & 0.483000 & 0.278980390 & 0 \\
  \emph{up} &0 & 0.469120 & 0.291124891 & 0 \\
  \emph{c1} &0 & 0.509000 & 0.254624859 & 0 \\
  \hline
 \end{tabular}
 \end{center}
\end{table}

We integrated the orbits until 1000 time units (t.u.), with a time step of 0.05
t.u.; the output to screen was enabled. The values of the indicators were dumped
to files every 20 time steps. The IDVs were chosen at random and
orthonormalized. Column H1 of Table \ref{tiempos} shows the outcome; as can be
seen comparing $T_1$ (individual CIs) with $T_2$ (simultaneous CIs), a reduction
of about 40 per cent in time was obtained by computing all the indicators at the
same time. Taking into account that GALI$_2$ is the same as SALI (though they
are computed through different algorithms), we repeated the above experiment but
without the computation of the SALI, which in this case is superfluous. The
resulting times ($T_1=8.0$, $T_2=5.1$) gave a reduction of 36 per cent.

\begin{table}
 \caption{CPU times of the code. $T_1$ stands for the sum of all the previous rows; $T_2$ is the time elapsed when all the indicators were computed at the same time. All times in seconds, rounded to the first decimal.}
 \label{tiempos}
 \begin{center}
 \begin{tabular}{lrrrrr}
  \hline
  & H1 &  H2 & H3 & N1 & N2\\
  \hline
  orbit & 0.4 &  ---  &     --- &   --- &   --- \\
  LI    & 1.5 &  36.0 &  4534.9 & 137.2 &   --- \\
  SALI  & 0.9 &  21.8 &  2750.1 &  44.5 &   --- \\ 
  GALIs & 2.5 &  64.5 &  8061.5 & 135.3 & 135.3 \\
  SD    & 0.9 &  22.5 &  2834.6 &  44.6 &   --- \\ 
  RLI   & 1.2 &  29.7 &  3753.4 &  50.9 &  50.9 \\
  MEGNO & 0.8 &  19.7 &  2473.2 &  32.8 &  32.8 \\ 
  FLI   & 0.7 &  15.4 &  1947.2 &  28.0 &  28.0 \\ 
  \hline
  $T_1$ & 8.9 & 209.6 & 26355.4 & 473.4 & 247.1 \\
  $T_2$ & 5.2 & 126.0 & 15770.5 & 370.5 & 204.6 \\
  \hline
 \end{tabular}
 \end{center}
\end{table}

The second set, dubbed H2, was built with 1000 i.c. taken from the region $x=0$,
$y\in [-0.1,0.1]$, $p_y=0$, with the energy fixed at $E=0.118$. The integration
time was 100 t.u., the time step was set to 0.05 t.u., the IDVs were chosen as
in experiment H1, and the output to screen was disabled. The indicators were
dumped only at their last values. The orbits were not dumped to file, in the
understanding that when there are lots of i.c., the user probably won't focus in
studying the aspect of each orbit in the configuration space. Column H2 of Table
\ref{tiempos} shows the elapsed times. Again, a reduction of about 40 per cent
in time was obtained. We repeated the experiment without the SALI, obtaining
$T_1=188.2$ and $T_2=125.8$, i.e., a reduction of 33 per cent.

The third set, dubbed H3, consisted of 125,751 i.c. from the region $x=0$, $y\in
[-0.1,0.1]$, $p_y=\in [-0.05,0.05]$, with the energy fixed at $E=0.118$. The
parameters were the same as in experiment H2. As can be seen from column H3 of
Table \ref{tiempos}, in this longer experiment the time saved was again about 40
percent. Without the SALI, we obtained $T_1=23604.7$ and $T_2=15734.1$, again, a
reduction of 33 per cent.

In order to show some actual input/output of the code, we now compute for the
H\'enon-Heiles potential a subset of CIs (the MEGNO, the LI, the RLI, the
GALI$_3$ and the OFLI) using the i.c. of experiment H1, plus a second chaotic
orbit lying in a large chaotic sea, dubbed {\it c2}, with i.c. $x=0$, $y=0.56$,
$p_y=0.112$ and $E=0.118$ \citep{CGS03}.  The  integration time was 15,000 t.u.
for each orbit, whereas the total cpu time was 55.5 s.

\begin{figure}
\epsfxsize=\hsize \epsfbox{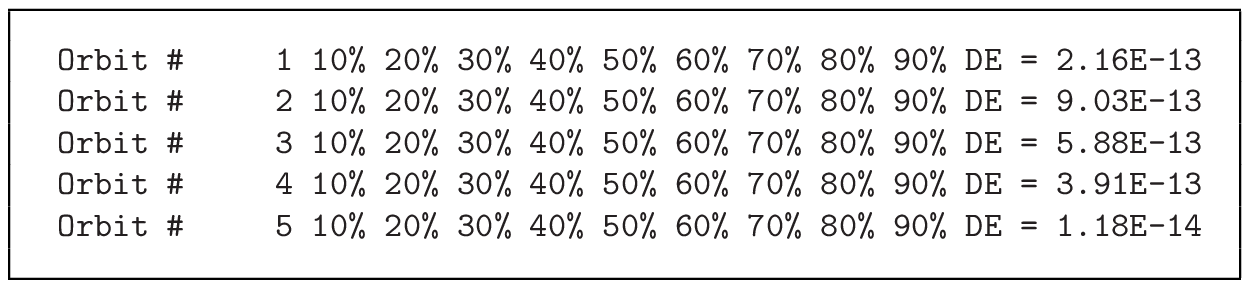}
\caption{Typical output to screen: orbit number, progress of the integration
(with percentages appearing on the screen as the integration progresses), and
conservation of the energy for each orbit.}
\label{out1}
\end{figure}

If the flag for visual control of the processing is enabled, the progress of the
computation and the energy conservation of each orbit should appear on the
screen, as is shown in Fig. \ref{out1}.

\begin{figure}
\epsfxsize=\hsize \epsfbox{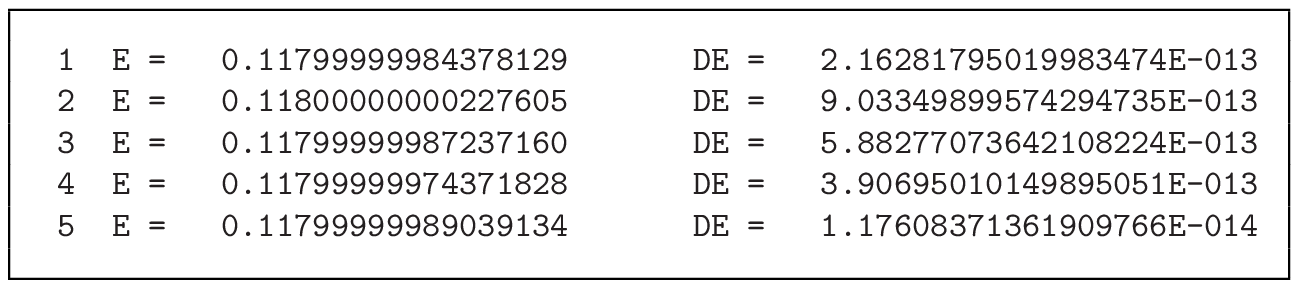}
\caption{Typical output to file {\tt *.ene}: orbit number, energy
and conservation of the energy for each orbit.}
\label{outene}
\end{figure}

One of the input parameters the user should give is an alphanumeric
prefix for the output files. Choosing for example {\tt hh} for the prefix of the
present run, the output files would be {\tt hh.ene}, {\tt hh.megno}, {\tt
hh.rli}, {\tt hh.gali} and {\tt hh.fli}. The first one will contain the energy
and its conservation for each orbit, as shown in Fig. \ref{outene}.

\begin{figure}
\epsfxsize=\hsize \epsfbox{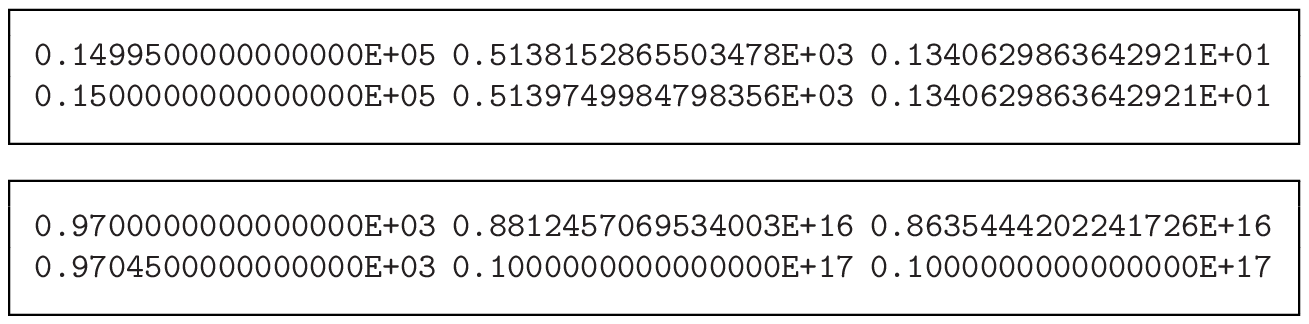}
\caption{Typical output to file {\tt *.fli} (only the two last rows
for the first and last orbits are shown): time, FLI and OFLI.}
\label{outfli}
\end{figure}

The files {\tt hh.megno}, {\tt hh.rli}, {\tt hh.gali} and {\tt hh.fli} will
contain the values of the CIs. Fig. \ref{outfli} shows the last two lines of the
first  and last orbits on the file {\tt hh.fli} of our example. It can be seen
that the first orbit reached the end of the integration (i.e., the FLI didn't
saturate), whereas the FLI of the last one saturated at $t=970.45$ t.u., from
which time the integration of this CI stopped. 

\begin{figure}
\includegraphics[width=0.5\textwidth]{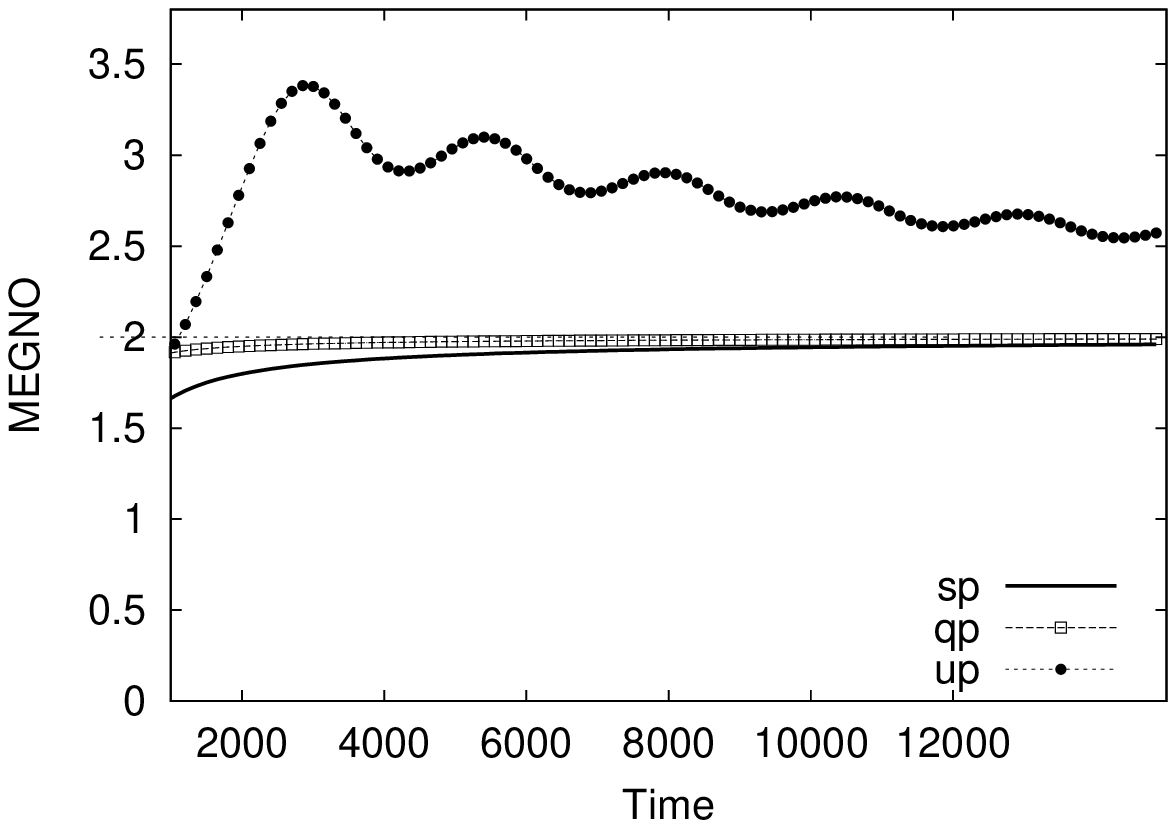}
\includegraphics[width=0.5\textwidth]{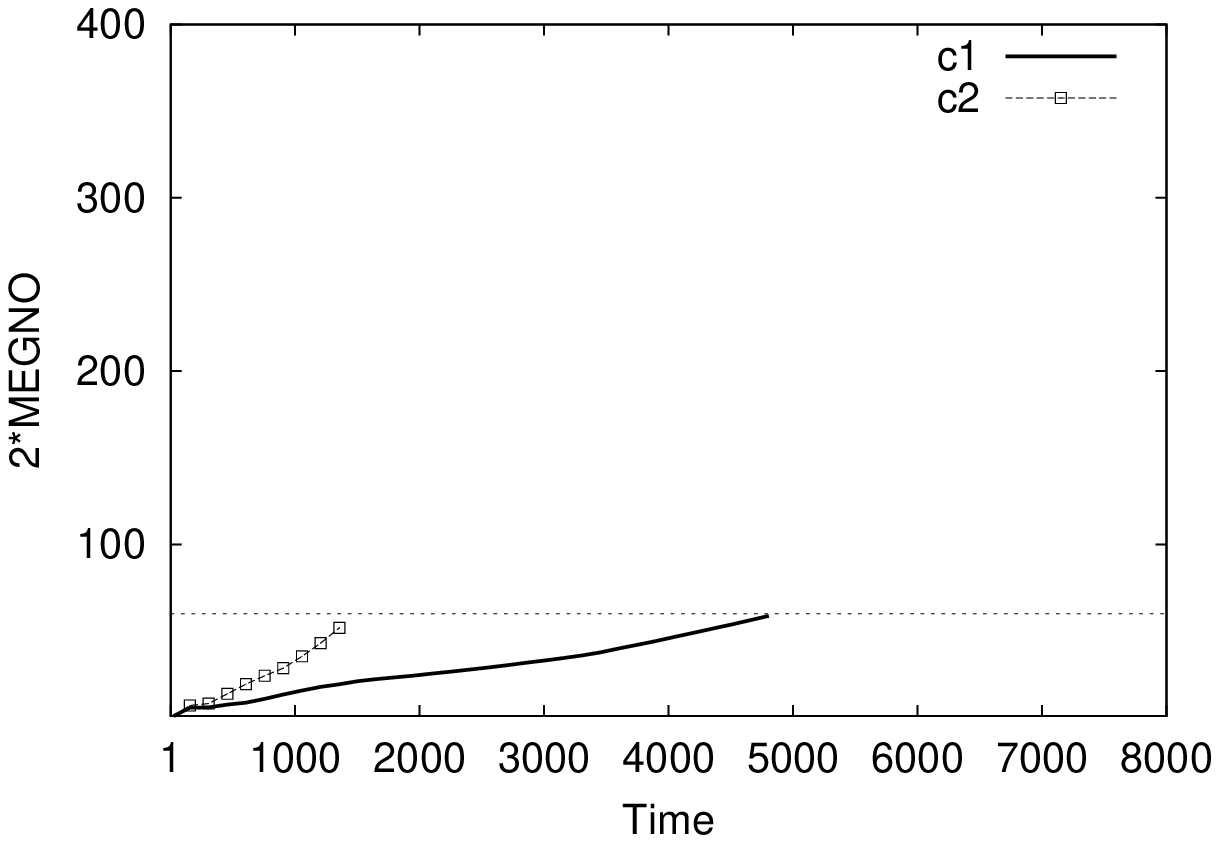}

\includegraphics[width=0.5\textwidth]{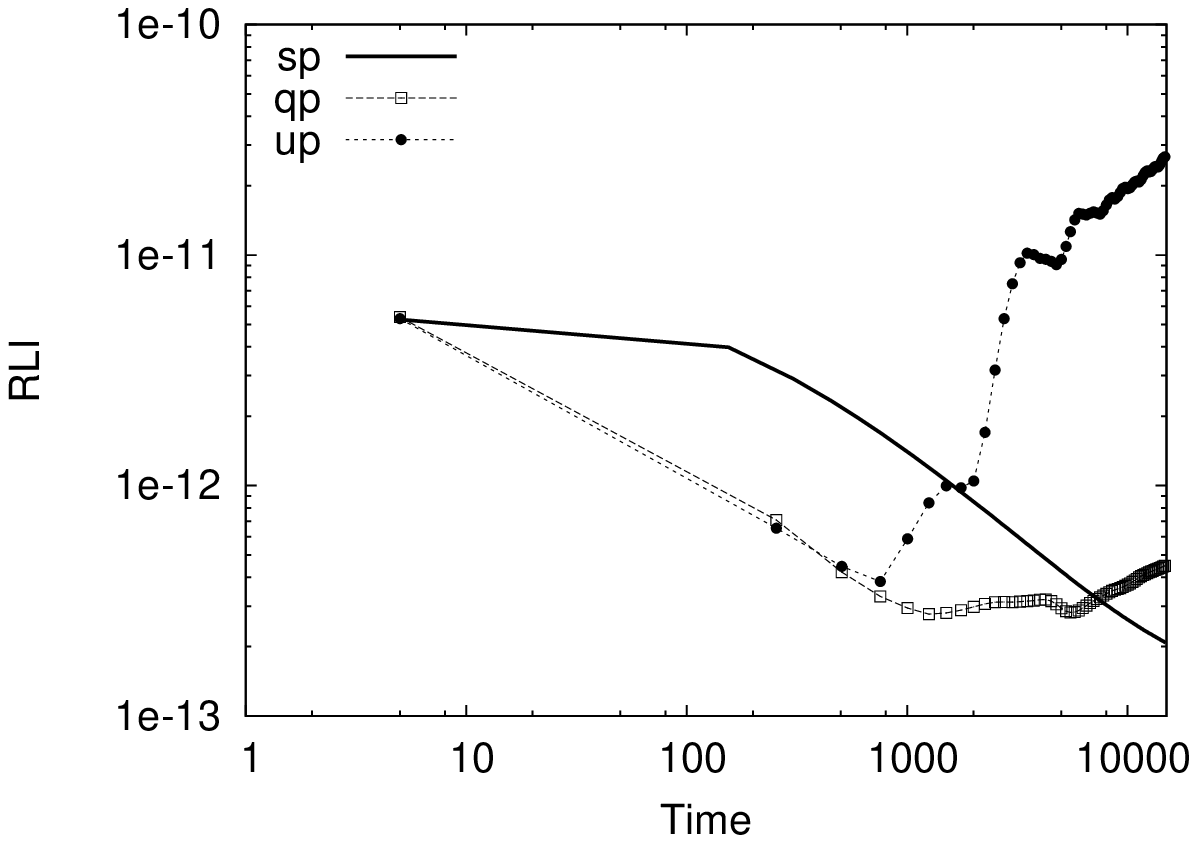}
\includegraphics[width=0.5\textwidth]{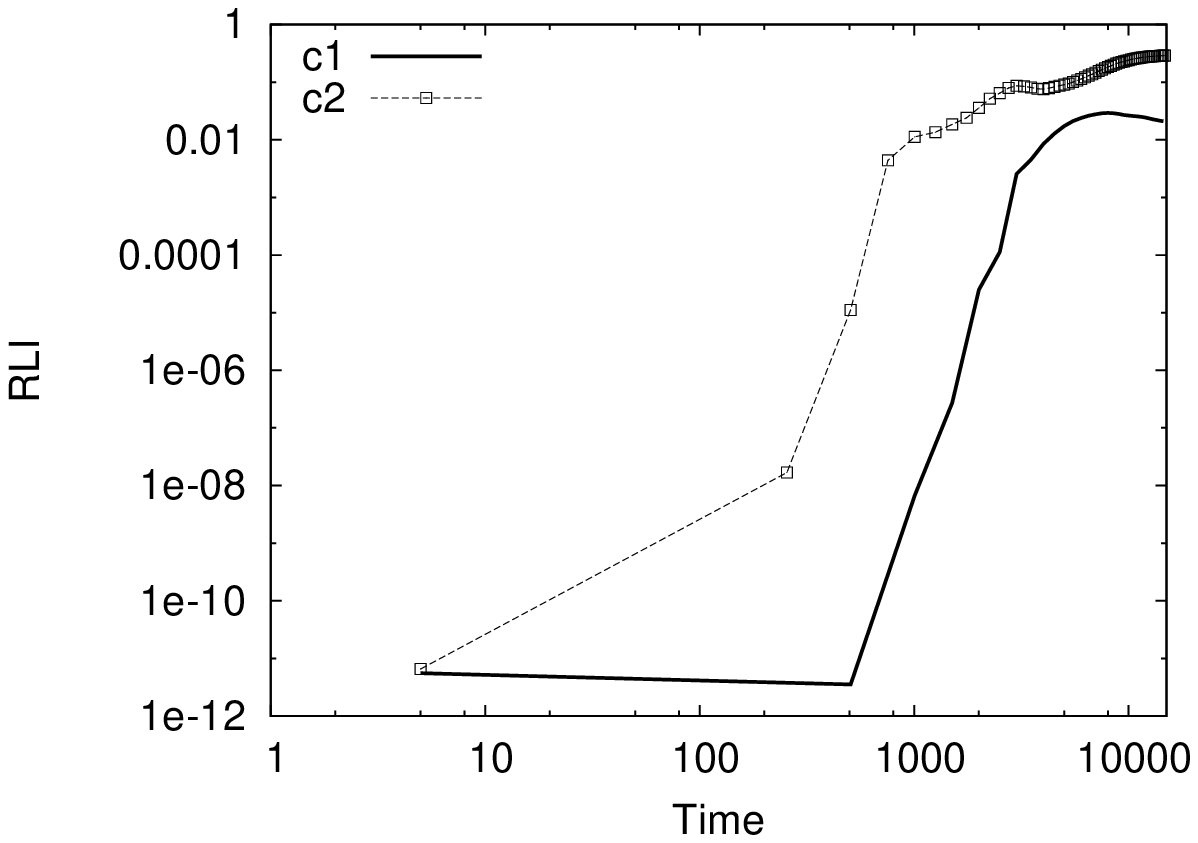}
\caption{Examples of the time evolution for the MEGNO (top panels)
and the RLI (bottom panels).}
\label{hh-1}
\end{figure}

\begin{figure}
\includegraphics[width=0.5\textwidth]{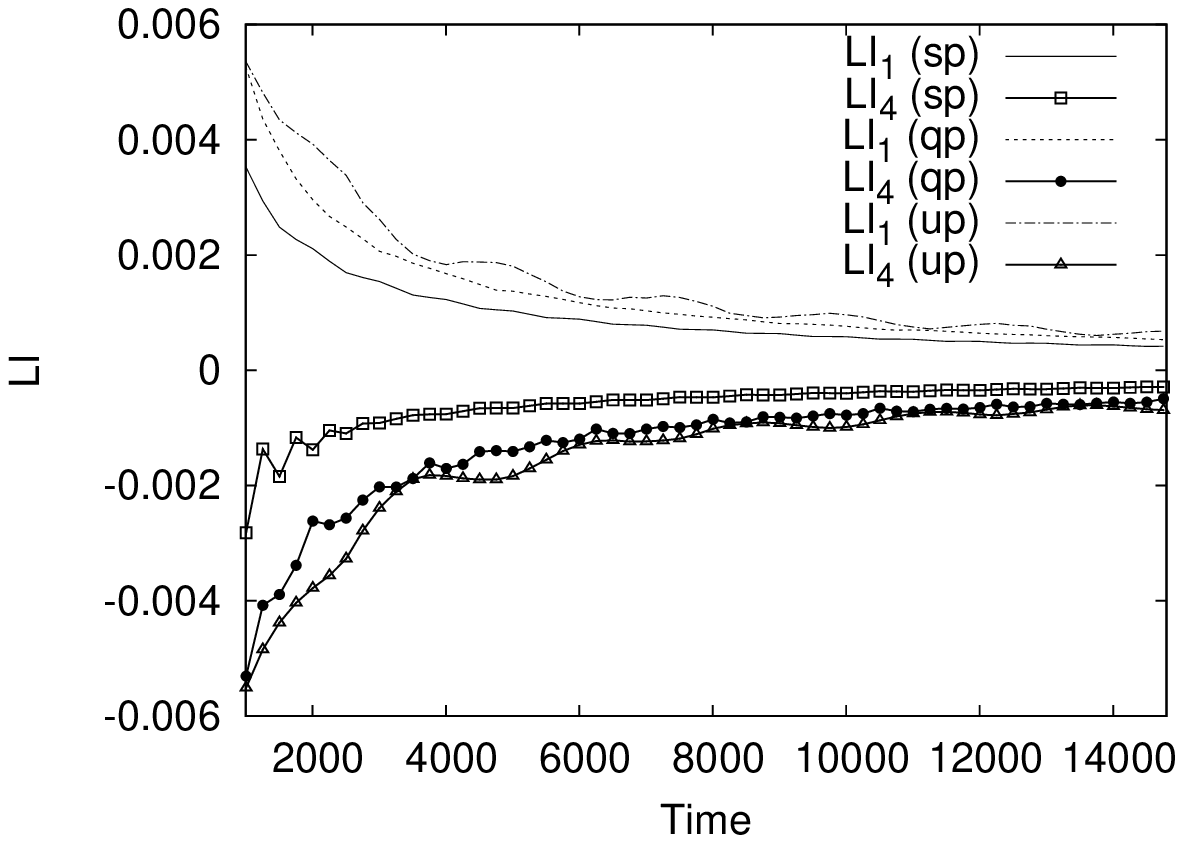}
\includegraphics[width=0.5\textwidth]{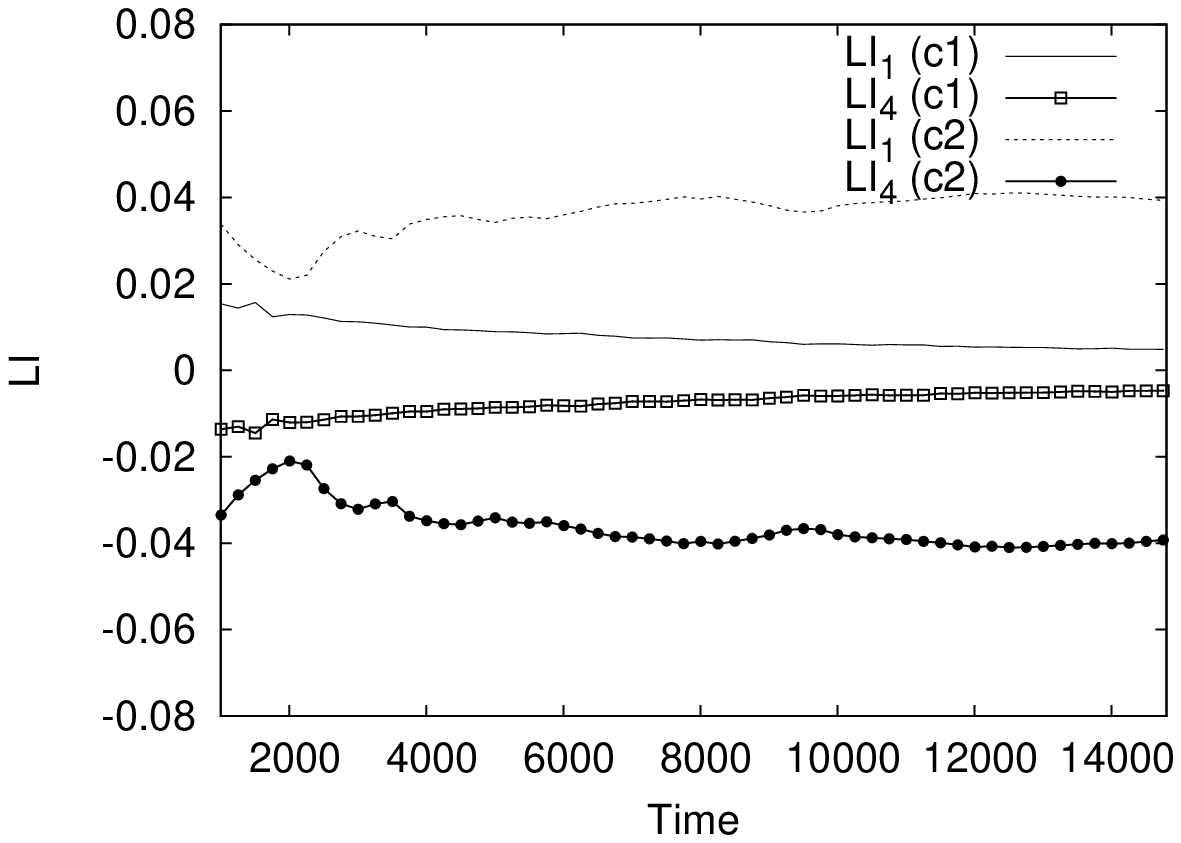}

\includegraphics[width=0.5\textwidth]{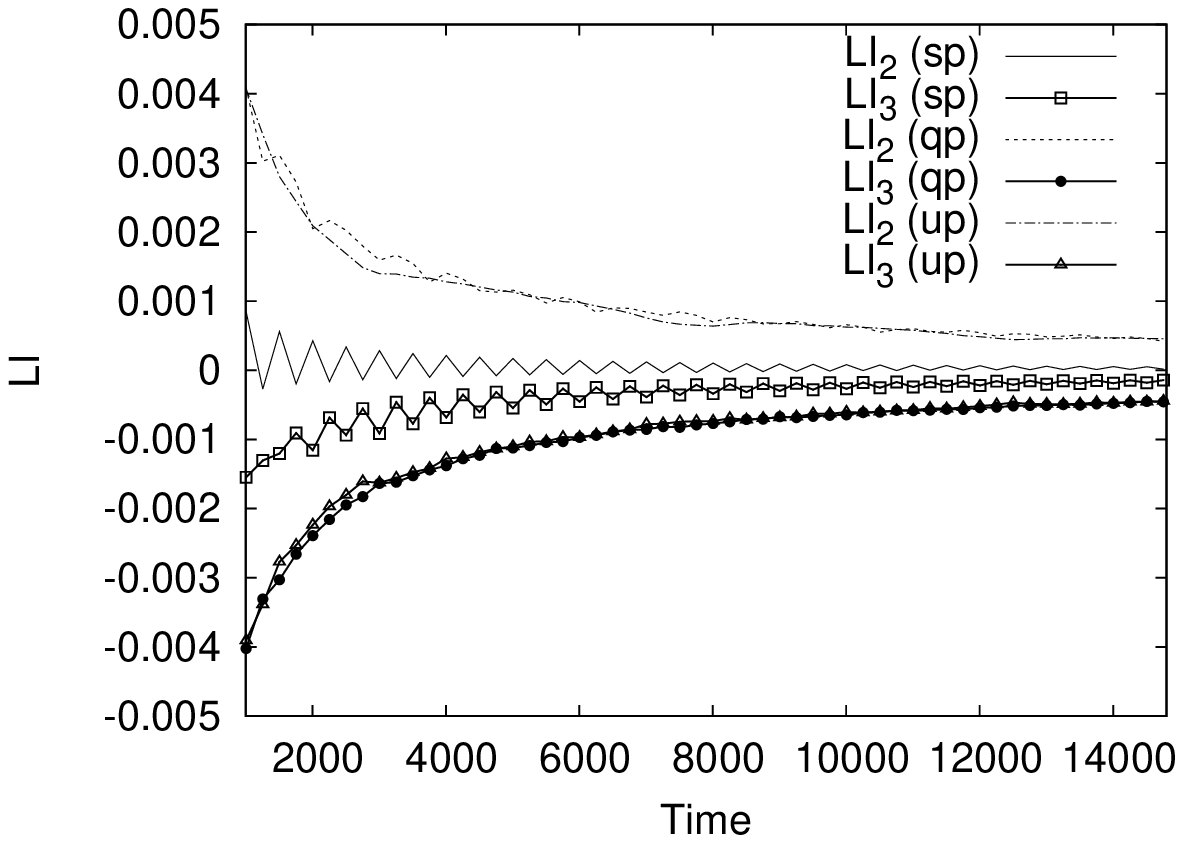}
\includegraphics[width=0.5\textwidth]{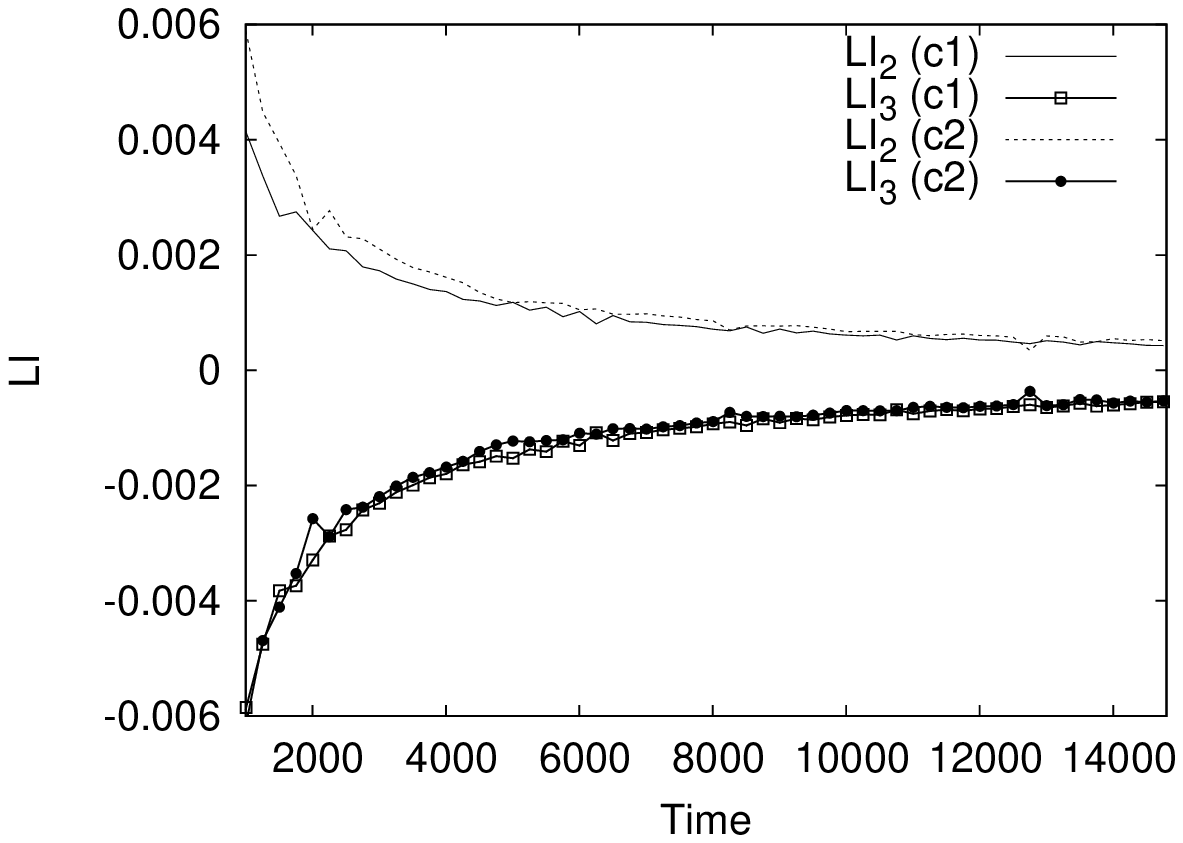}

\caption{Examples of the time evolution of the LI$_1$ and LI$_4$
(top panels), and of the LI$_2$ and LI$_3$ (bottom panels)} 
\label{hh-1LI}
\end{figure}

\begin{figure}
\includegraphics[width=0.5\textwidth]{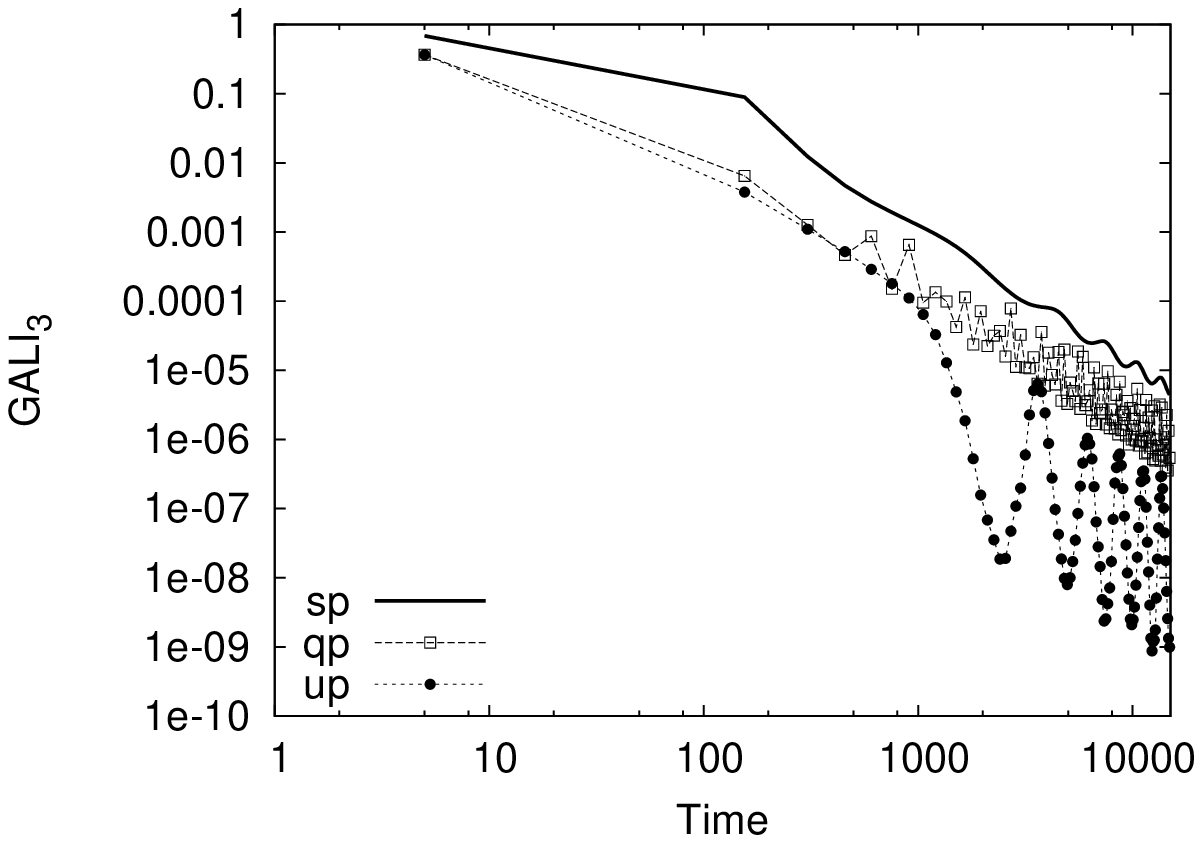}
\includegraphics[width=0.5\textwidth]{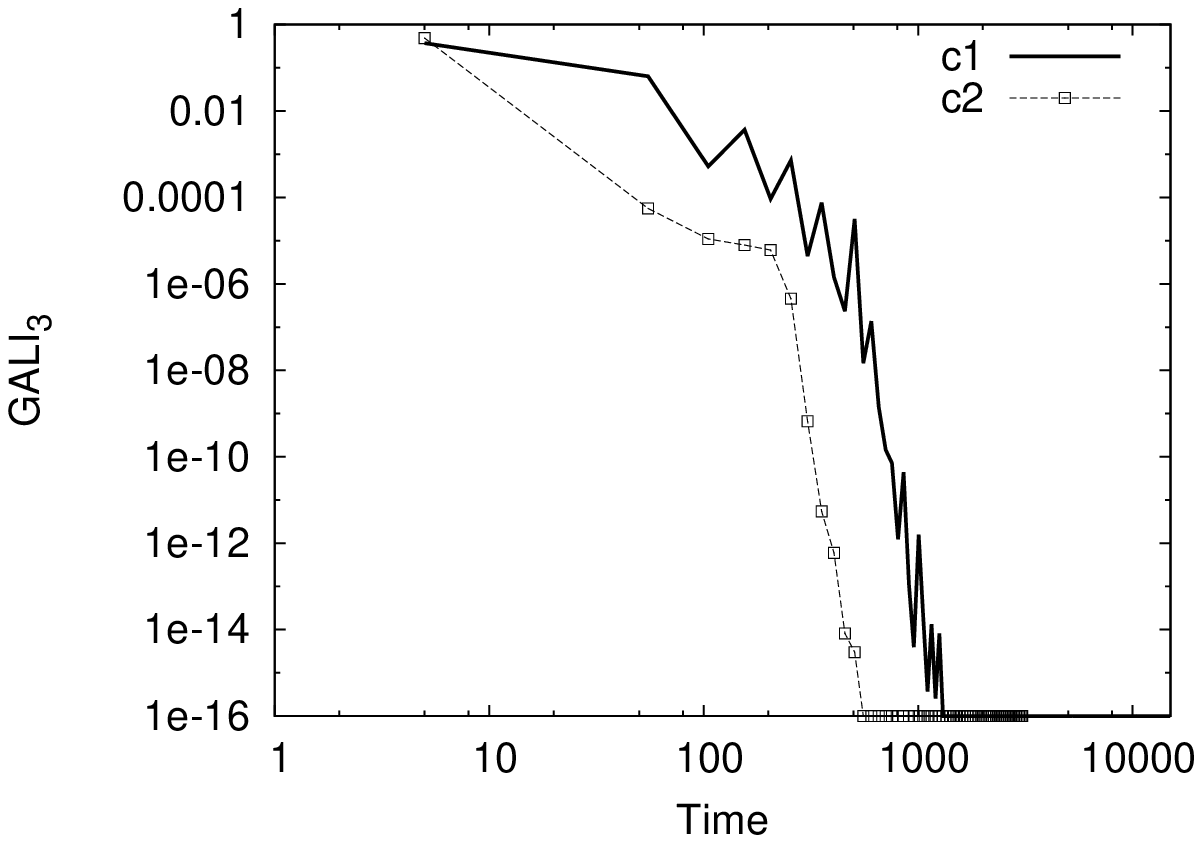}

\includegraphics[width=0.5\textwidth]{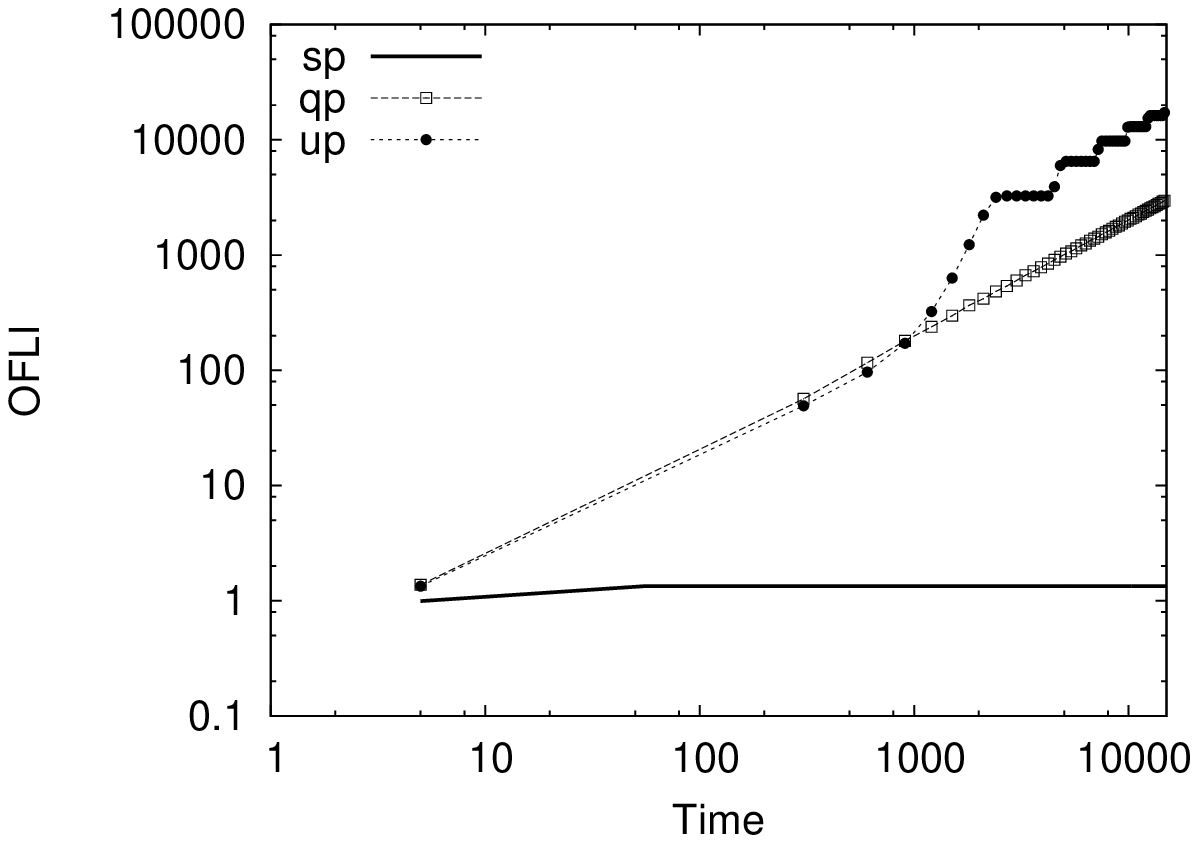}
\includegraphics[width=0.5\textwidth]{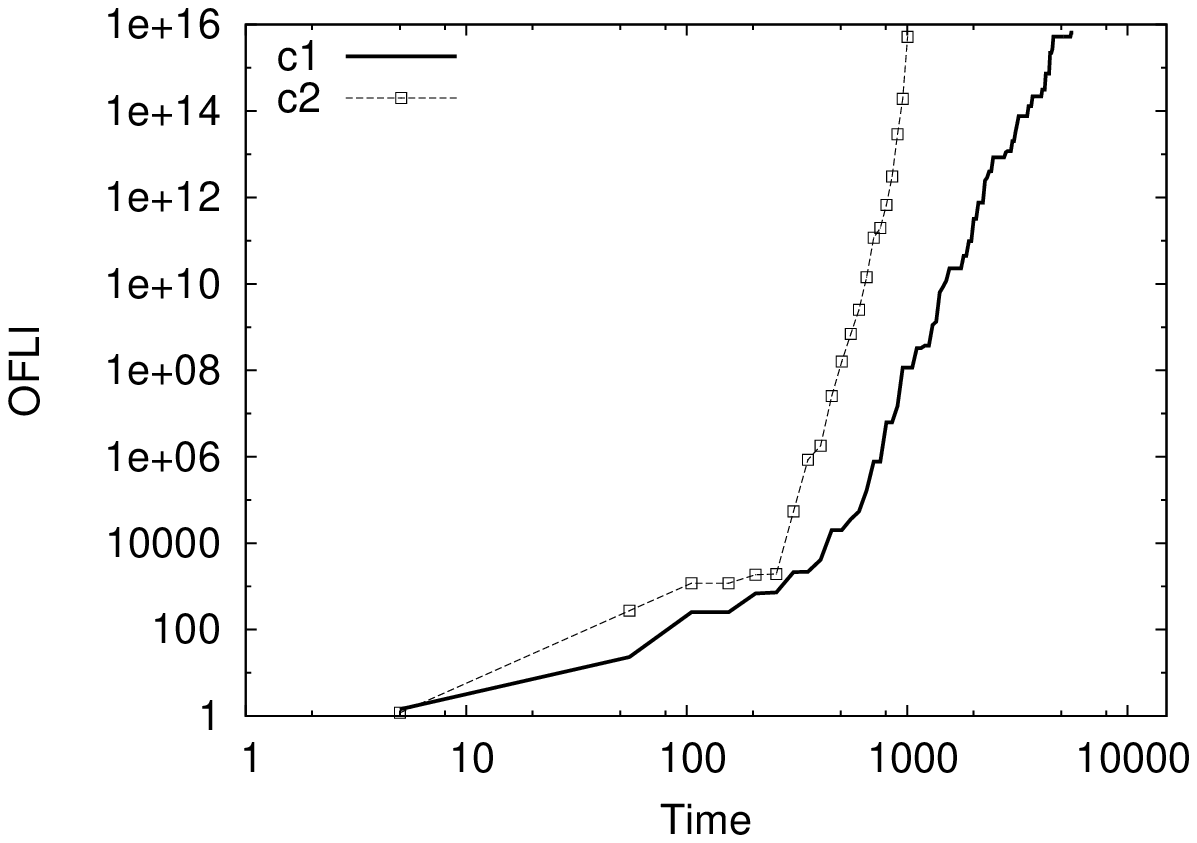}
\caption{Examples of the time evolution for the GALI$_3$ (top
panels) and the OFLI (bottom panels).}
\label{hh-2}
\end{figure}

Figs. \ref{hh-1}, \ref{hh-1LI} and \ref{hh-2} show the resulting
CIs for the five orbits considered. The top left panel of Fig. \ref{hh-1} shows
the behaviour of the MEGNO for the three regular orbits. We can see that the
orbit {\it sp} tends to $2$ from below and the orbit {\it up} tends to $2$ from
above, as expected from the dependence of the MEGNO on the stability of the
orbit \citep{CGS03}. On the other hand, the orbit {\it qp} quickly increases to
$2$, again as expected. The top right panel of Fig. \ref{hh-1} shows the
behaviour of the MEGNO for the chaotic orbits. It is clearly seen that, once the
MEGNO reached its saturation value, its computation was stopped. These plots
should be compared with those of fig. 1(d) and (c) of \citet{CGS03}. The bottom
panels of Fig. \ref{hh-1} show the RLI for the five orbits of the sample. As
expected, the RLI maintains very small values for regular orbits, and reaches
relatively high values for chaotic ones.

Fig. \ref{hh-1LI} shows the evolution of the different LIs for the
three regular orbits (left panels) and for the two chaotic orbits (right
panels). The top panels show the evolution of LI$_1$ and LI$_4$, whereas the
bottom ones show the values of LI$_2$ and LI$_3$. These figures should be
compared with figures 1(d) and 1(e) of \citet{CGS03}. In all cases, the LI
behaves as expected. Though the LI$_1$ curve for the {\it c1} orbit seems to be
tending to zero, it is really tending to a constant value greater than zero.
This can be clearly seen in a logarithmic scale; we refrained to use that scale
in order to be able to show all the LIs, both positive and negative.

The top panels of Fig. \ref{hh-2} show the GALI$_3$ for our regular (left) and
chaotic (right) orbits. The indicator behaves, again, as expected: a polynomial
decrease in the case of regular orbits, and an exponential decrease in the
chaotic case. The bottom panels of Fig. \ref{hh-2}  present the OFLI for the
five previous orbits. On the left panel, the CI shows a linear increment with
time for  the orbits {\it qp} and {\it up}, and a constant value for the orbit
{\it sp}. The behaviour for the orbit {\it sp} reflects its proximity to a 
periodic orbit. For chaotic orbits, instead, the OFLI grows exponentially fast.
It is seen that this CI reached the saturation value ($10^{16}$) for both
chaotic orbits before the end of the integration, and their computation was
therefore stopped.    

Additionally, we have reproduced many other plots from papers in which different
potentials and/or CIs were studied. We refrain from showing these plots because
they are identical with the original ones. Suffice to say that we didn't find
any instance of a different result from those of the literature.

\subsection{The triaxial NFW potential}

We also probed our code with a 3D potential, which is a triaxial extension of
the NFW profile \citep{VWHS08}:
\begin{equation}
\Phi_{\rm N}=-\frac{A}{r_p}\ln \left( 1+\frac{r_p}{r_s}\right),
\end{equation}
where $A$ and $r_s$ are constants, and
\begin{equation}
r_p=\frac{(r_s+r)r_e}{r_s+r_e},
\end{equation}
with
\begin{equation}
r_e=\sqrt{\left(\frac{x}{a}\right)^2 + \left(\frac{y}{b}\right)^2 +
\left(\frac{z}{c}\right)^2},
\end{equation}
$a$, $b$, $c$ constants, and $r=\sqrt{x^2+y^2+z^2}$. The values of the constants
we used are listed in Table \ref{const}. In this experiment, dubbed N1, in which
we integrated 140 orbits, the total integration time was 13 t.u., which with our
choice of constants corresponds to 13  Gyr. The time step was 0.005  t.u., and
the rest of parameters were as in experiment H1. In column N1 of Table
\ref{tiempos} we list the CPU times of this experiment. The  saved time was, in
this case, around 22 per cent. Avoiding the SALI as in the former cases, the 
runs lasted $T_1=428.9$ t.u. and $T_2=369.6$ t.u., a 14 per cent of saved time.

We repeated this last experiment but with only the GALIs, the RLI, the MEGNO and
the OFLI enabled (experiment N2). Since the SALI and the DS share equations with
the GALIs, we expect that the saved time will be less than before. This is
confirmed in column N2 of Table \ref{tiempos}, where we can see now that the
percentage gained is about 17 per cent.

\begin{table}
 \caption{Constants used for the $\Phi_{\rm N}$ potential.}
 \label{const}
 \begin{center}
 \begin{tabular}{rr@{.}l}
  \hline
  $A$ & 4158670&1856267899 \\
  $r_s$ & 19&044494521343964 \\
  $a$ & 1&3258820840000000 \\
  $b$ & 0&86264540200000000 \\
  $c$ & 0&70560584600000000 \\
  \hline
 \end{tabular}
 \end{center}
\end{table}

\begin{figure}
\epsfxsize=\hsize \epsfbox{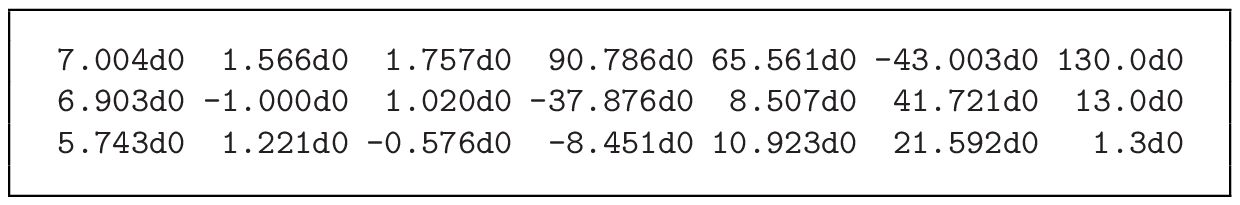}
\caption{Initial conditions for experiment N3. For each orbit
(row) the given data are: $x$, $y$, $z$,
$\dot x$, $\dot y$, $\dot z$ and the desired time of integration.}
\label{innfw}
\end{figure}

Finally, we tested the implementation of the CIs that have not been used in  the
last experiment. In this new run, N3, we computed the LIs, the SALI and the FLI
of three orbits for different intervals of integration. The values of the IDVs
were taken random and orthonormal. The i.c. are listed in Fig. \ref{innfw}, just
like it would appear in the i.c. file. Notice that, since we want the orbits to
be integrated for different time intervals, the latter may be inserted at the
end of each row of i.c.; in this way, the program automatically understands that
the time of integration should be taken from these numbers.

\begin{figure}
\includegraphics[width=0.5\textwidth]{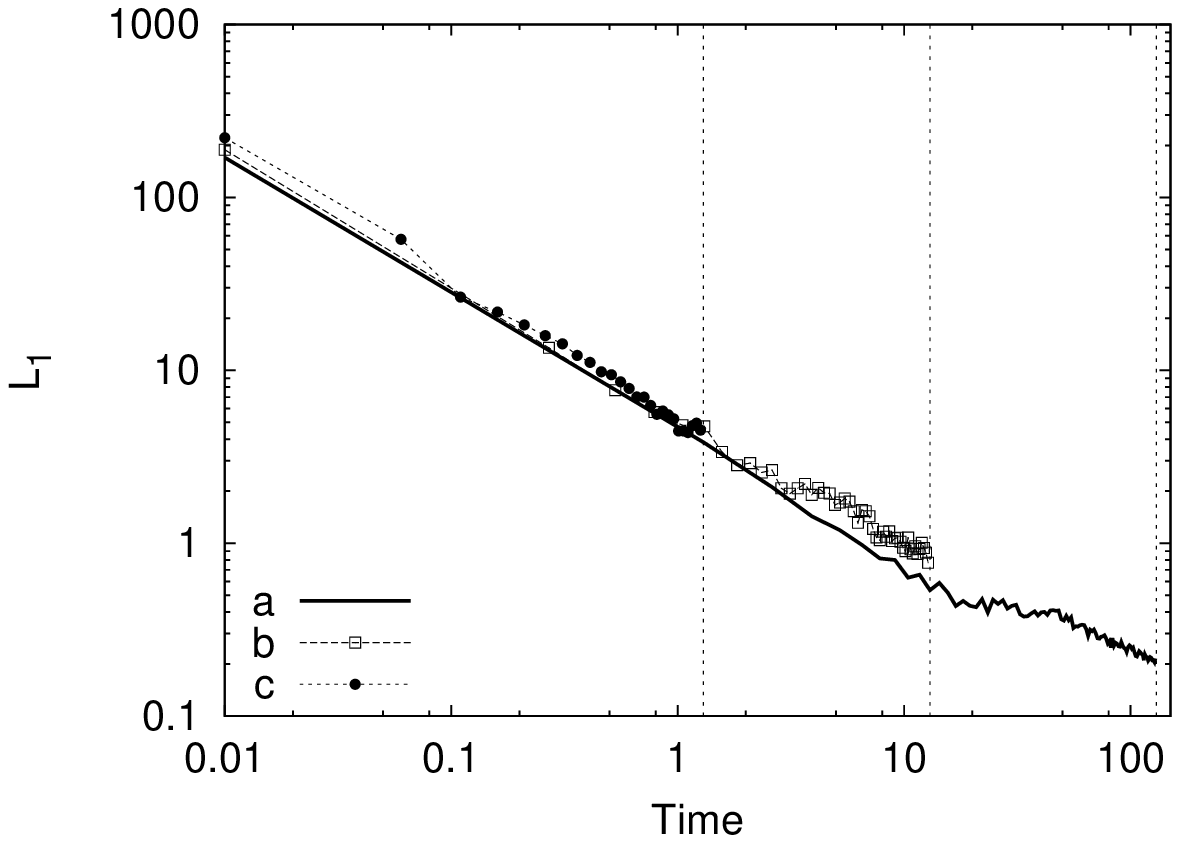}
\includegraphics[width=0.5\textwidth]{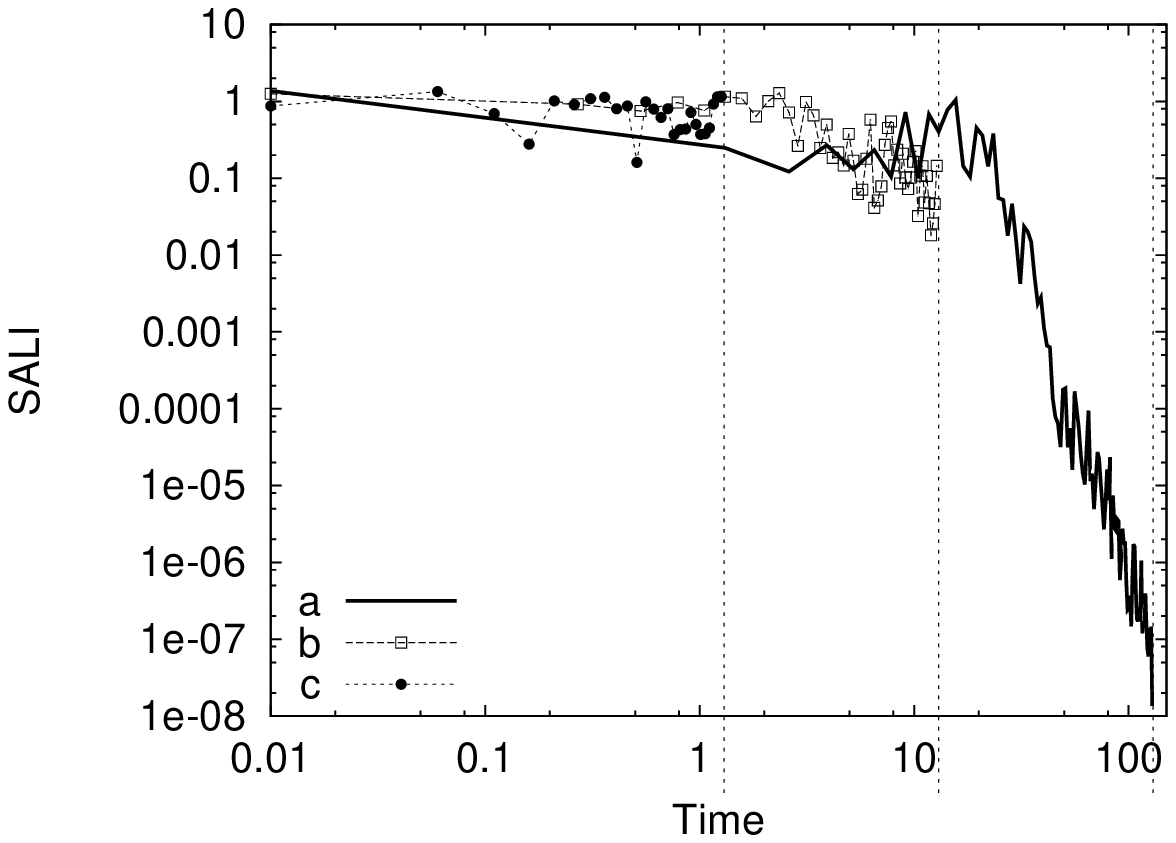}

\hfil\includegraphics[width=0.5\textwidth]{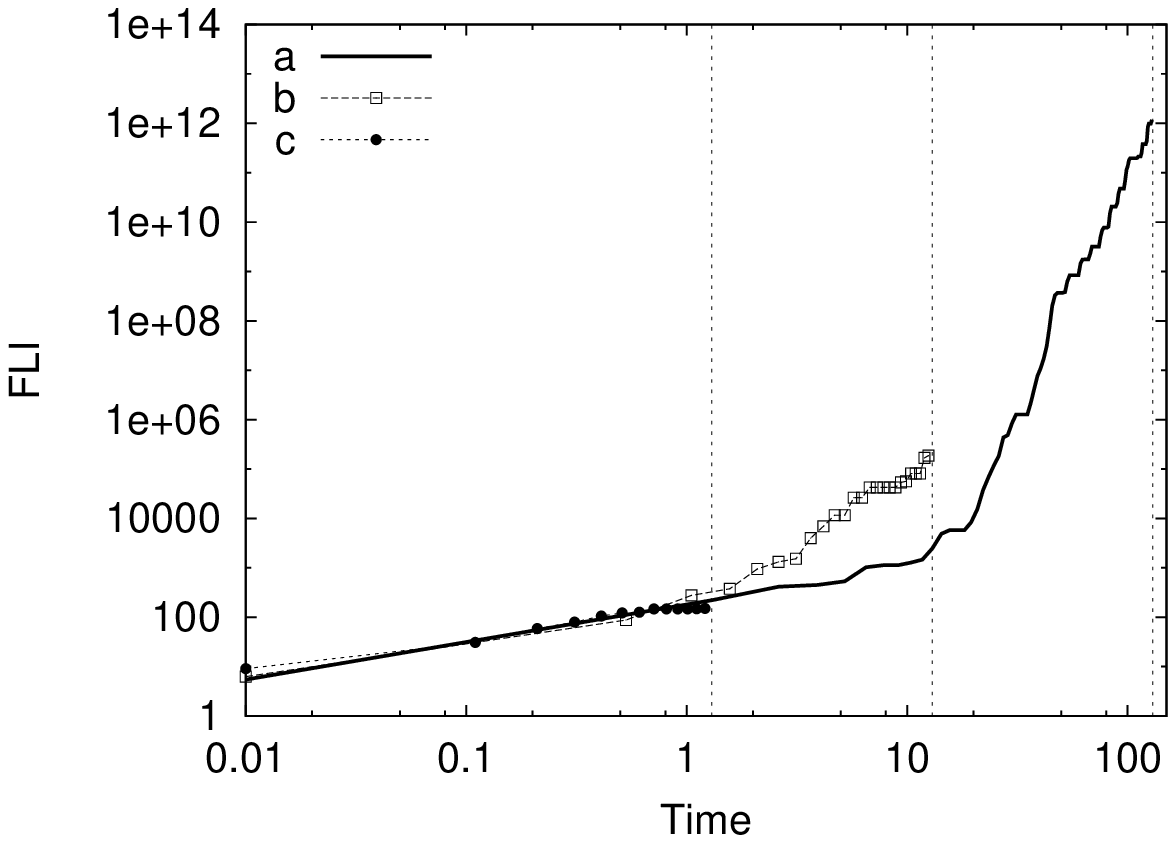}\hfil
\caption{Examples of the time evolution for the LI (top
left panel, labelled L$_1$), the SALI (top right panel) and the FLI (bottom
panel) for the 3 orbits of experiment N3.}
\label{nfw-1}
\end{figure}

Fig. \ref{nfw-1} shows the results. The vertical dashed lines mark the three
different times of integration used in the experiment. The top left panel shows
the LI: it decreases for the three orbits, thus identifying them as regular.
However, the SALI (top right panel) decreases exponentially fast by the end of
the time interval for orbit (a), which means that the latter is chaotic. The
SALI of the other two orbits, on the other hand, seem to oscillate around a
finite value, that is, they have the behaviour corresponding to regular motion.
The bottom panel shows the FLI. It grows exponentially for orbits (a) and (b),
implying that both orbits are chaotic. Therefore, it is clear that 13 and 1.3
t.u. (the integration times used for orbits (b) and (c), respectively) are not
enough to reliably classify them. Nevertheless, the FLI and the SALI prove to be
faster indicators than the LI, as many papers in the literature stand for 
\citep[e.g.,][]{S01,SESF04}.

We have also probed the {\tt LP-VIcode} with a variety of other astronomical
potentials, both 2D and 3D, playing with different combinations of indicators,
and obtained the expected results in all cases. For example, adding the
computation of the FLI to an experiment in which the MEGNO is already enabled,
does not involve any additional computational time. Or, in computing the LIs and
the GALIs at the same time, the only gain is that the orbit is computed only
once, due to the fact that these indicators don't share any variational
equations.

\section{Conclusions}

The main goal of the {\tt LP-VIcode} is to cluster in a single, easy-to-use tool
the plethora of CIs that are nowadays in the literature. The starting point is
the present code, which is ready to use, except for the routines that compute
the potential, the accelerations and the variational equations of the system,
which should be provided by the user. The program can handle any number of
dimensions, and is not limited to stellar or planetary systems, but it works
with any dynamical system in which the abovementioned equations can be written
down. The code is optimized to achieve the maximum speed, given a set of CIs to
compute.

We expect researchers to collaborate with their own methods in developing newer
versions of the code containing larger CIs' libraries. Also, we are open to
people who might be interested in making the code more user-friendly, for
example, changing the present command-driven interface to a menu-driven
interface, etc.

The code, the User's Guide and complete examples are available at\break
{\tt www.fcaglp.unlp.edu.ar/LP-VIcode/}.

\section{Acknowledgements}

The authors wish to express their deep gratitude to Pablo Cincotta and Claudia
Giordano for their fundamental contributions to the alpha version of the {\tt
LP-VIcode}. This research was supported with grants from the Universidad
Nacional de La Plata, Argentina, the Consejo Nacional de Investigaciones
Cient\'\i ficas y T\'ecnicas de la Rep\'ublica Argentina, and the Agencia
Nacional de Promoci\'on Cient\'\i fica y Tecnol\'ogica  de la Rep\'ublica
Argentina.

\bibliographystyle{model2-names}
\bibliography{biblio.bib}

\begin{thebibliography}{43}
\expandafter\ifx\csname natexlab\endcsname\relax\def\natexlab#1{#1}\fi
\expandafter\ifx\csname url\endcsname\relax
  \def\url#1{\texttt{#1}}\fi
\expandafter\ifx\csname urlprefix\endcsname\relax\def\urlprefix{URL }\fi
\providecommand{\eprint}[2][]{\url{#2}}
\providecommand{\bibinfo}[2]{#2}
\ifx\xfnm\relax \def\xfnm[#1]{\unskip,\space#1}\fi
%Type = Article
\bibitem[{{Benettin} et~al.(1980a){Benettin}, {Galgani}, {Giorgilli} and
  {Strelcyn}}]{BGGS80a}
\bibinfo{author}{{Benettin}, G.}, \bibinfo{author}{{Galgani}, L.},
  \bibinfo{author}{{Giorgilli}, A.}, \bibinfo{author}{{Strelcyn}, J.M.},
  \bibinfo{year}{1980}a.
\newblock \bibinfo{title}{Lyapunov characteristic exponents for smooth
  dynamical systems and for hamiltonian systems; a method for computing all of
  them. part 1: Theory}.
\newblock \bibinfo{journal}{Meccanica} \bibinfo{volume}{15},
  \bibinfo{pages}{9--20}.
%Type = Article
\bibitem[{{Benettin} et~al.(1980b){Benettin}, {Galgani}, {Giorgilli} and
  {Strelcyn}}]{BGGS80b}
\bibinfo{author}{{Benettin}, G.}, \bibinfo{author}{{Galgani}, L.},
  \bibinfo{author}{{Giorgilli}, A.}, \bibinfo{author}{{Strelcyn}, J.M.},
  \bibinfo{year}{1980}b.
\newblock \bibinfo{title}{Lyapunov characteristic exponents for smooth
  dynamical systems and for hamiltonian systems; a method for computing all of
  them. part 2: Numerical application}.
\newblock \bibinfo{journal}{Meccanica} \bibinfo{volume}{15},
  \bibinfo{pages}{21--30}.
%Type = Article
\bibitem[{{Benettin} et~al.(1976){Benettin}, {Galgani} and {Strelcyn}}]{BGS76}
\bibinfo{author}{{Benettin}, G.}, \bibinfo{author}{{Galgani}, L.},
  \bibinfo{author}{{Strelcyn}, J.M.}, \bibinfo{year}{1976}.
\newblock \bibinfo{title}{{Kolmogorov entropy and numerical experiments}}.
\newblock \bibinfo{journal}{\pra} \bibinfo{volume}{14},
  \bibinfo{pages}{2338--2345}.
%Type = Article
\bibitem[{{Binney} and {Spergel}(1982)}]{BS82}
\bibinfo{author}{{Binney}, J.}, \bibinfo{author}{{Spergel}, D.},
  \bibinfo{year}{1982}.
\newblock \bibinfo{title}{{Spectral stellar dynamics}}.
\newblock \bibinfo{journal}{\apj} \bibinfo{volume}{252},
  \bibinfo{pages}{308--321}.
%Type = Book
\bibitem[{{Binney} and {Tremaine}(2008)}]{BT08}
\bibinfo{author}{{Binney}, J.}, \bibinfo{author}{{Tremaine}, S.},
  \bibinfo{year}{2008}.
\newblock \bibinfo{title}{Galactic Dynamics: (Second Edition)}.
\newblock Princeton Series in Astrophysics, \bibinfo{publisher}{Princeton
  University Press}.
%Type = Article
\bibitem[{{Carpintero} and {Aguilar}(1998)}]{CA98}
\bibinfo{author}{{Carpintero}, D.D.}, \bibinfo{author}{{Aguilar}, L.A.},
  \bibinfo{year}{1998}.
\newblock \bibinfo{title}{{Orbit classification in arbitrary 2D and 3D
  potentials}}.
\newblock \bibinfo{journal}{\mnras} \bibinfo{volume}{298},
  \bibinfo{pages}{1--21}.
%Type = Article
\bibitem[{{Cincotta} et~al.(2003){Cincotta}, {Giordano} and {Sim{\'o}}}]{CGS03}
\bibinfo{author}{{Cincotta}, P.M.}, \bibinfo{author}{{Giordano}, C.M.},
  \bibinfo{author}{{Sim{\'o}}, C.}, \bibinfo{year}{2003}.
\newblock \bibinfo{title}{{Phase space structure of multi-dimensional systems
  by means of the mean exponential growth factor of nearby orbits}}.
\newblock \bibinfo{journal}{Physica D Nonlinear Phenomena}
  \bibinfo{volume}{182}, \bibinfo{pages}{151--178}.
%Type = Article
\bibitem[{{Cincotta} and {Sim{\'o}}(2000)}]{CS00}
\bibinfo{author}{{Cincotta}, P.M.}, \bibinfo{author}{{Sim{\'o}}, C.},
  \bibinfo{year}{2000}.
\newblock \bibinfo{title}{{Simple tools to study global dynamics in
  non-axisymmetric galactic potentials - I}}.
\newblock \bibinfo{journal}{\aaps} \bibinfo{volume}{147},
  \bibinfo{pages}{205--228}.
%Type = Article
\bibitem[{{Contopoulos} and {Voglis}(1996)}]{CV96}
\bibinfo{author}{{Contopoulos}, G.}, \bibinfo{author}{{Voglis}, N.},
  \bibinfo{year}{1996}.
\newblock \bibinfo{title}{{Spectra of Stretching Numbers and Helicity Angles in
  Dynamical Systems}}.
\newblock \bibinfo{journal}{Celest. Mech. Dynam. Astron.} \bibinfo{volume}{64},
  \bibinfo{pages}{1--20}.
%Type = Inproceedings
\bibitem[{{Darriba} et~al.(2012a){Darriba}, {Maffione}, {Cincotta} and
  {Giordano}}]{DMCG12p}
\bibinfo{author}{{Darriba}, L.A.}, \bibinfo{author}{{Maffione}, N.P.},
  \bibinfo{author}{{Cincotta}, P.M.}, \bibinfo{author}{{Giordano}, C.M.},
  \bibinfo{year}{2012}a.
\newblock \bibinfo{title}{Chaos detection tools: the lp-vicode and its
  applications}, in: \bibinfo{editor}{P.~M.~{Cincotta}, C. M.~{Giordano}, C.E.}
  (Ed.), \bibinfo{booktitle}{3rd La Plata International School on Astronomy and
  Geophysics: Chaos, diffusion and non-integrability in Hamiltonian systems -
  Application to astronomy}, \bibinfo{publisher}{Universidad Nacional de La
  Plata - Asociaci\'on Argentina de Astronom\'\i a}. pp.
  \bibinfo{pages}{345--366}.
%Type = Article
\bibitem[{{Darriba} et~al.(2012b){Darriba}, {Maffione}, {Cincotta} and
  {Giordano}}]{DMCG12}
\bibinfo{author}{{Darriba}, L.A.}, \bibinfo{author}{{Maffione}, N.P.},
  \bibinfo{author}{{Cincotta}, P.M.}, \bibinfo{author}{{Giordano}, C.M.},
  \bibinfo{year}{2012}b.
\newblock \bibinfo{title}{Comparative study of variational chaos indicators and
  odes' numerical integrators}.
\newblock \bibinfo{journal}{International Journal of Bifurcation and Chaos}
  \bibinfo{volume}{22}, \bibinfo{pages}{1230033}.
%Type = Article
\bibitem[{{Deibel} et~al.(2011){Deibel}, {Valluri} and {Merritt}}]{DVM11}
\bibinfo{author}{{Deibel}, A.T.}, \bibinfo{author}{{Valluri}, M.},
  \bibinfo{author}{{Merritt}, D.}, \bibinfo{year}{2011}.
\newblock \bibinfo{title}{{The Orbital Structure of Triaxial Galaxies with
  Figure Rotation}}.
\newblock \bibinfo{journal}{\apj} \bibinfo{volume}{728}, \bibinfo{pages}{128}.
\newblock \eprint{1008.2753}.
%Type = Article
\bibitem[{{Eckmann} and {Ruelle}(1985)}]{ER85}
\bibinfo{author}{{Eckmann}, J.P.}, \bibinfo{author}{{Ruelle}, D.},
  \bibinfo{year}{1985}.
\newblock \bibinfo{title}{{Ergodic theory of chaos and strange attractors}}.
\newblock \bibinfo{journal}{Reviews of Modern Physics} \bibinfo{volume}{57},
  \bibinfo{pages}{617--656}.
%Type = Article
\bibitem[{{Fouchard} et~al.(2002){Fouchard}, {Lega}, {Froeschl{\'e}} and
  {Froeschl{\'e}}}]{FLFF02}
\bibinfo{author}{{Fouchard}, M.}, \bibinfo{author}{{Lega}, E.},
  \bibinfo{author}{{Froeschl{\'e}}, C.}, \bibinfo{author}{{Froeschl{\'e}}, C.},
  \bibinfo{year}{2002}.
\newblock \bibinfo{title}{{On the Relationship Between Fast Lyapunov Indicator
  and Periodic Orbits for Continuous Flows}}.
\newblock \bibinfo{journal}{Celest. Mech. Dynam. Astron.} \bibinfo{volume}{83},
  \bibinfo{pages}{205--222}.
%Type = Article
\bibitem[{{Froeschl{\'e}} et~al.(1997){Froeschl{\'e}}, {Gonczi} and
  {Lega}}]{FGL97}
\bibinfo{author}{{Froeschl{\'e}}, C.}, \bibinfo{author}{{Gonczi}, R.},
  \bibinfo{author}{{Lega}, E.}, \bibinfo{year}{1997}.
\newblock \bibinfo{title}{{The fast Lyapunov indicator: a simple tool to detect
  weak chaos. Application to the structure of the main asteroidal belt}}.
\newblock \bibinfo{journal}{\planss} \bibinfo{volume}{45},
  \bibinfo{pages}{881--886}.
%Type = Article
\bibitem[{{H\'enon} and {Heiles}(1964)}]{HH64}
\bibinfo{author}{{H\'enon}, M.}, \bibinfo{author}{{Heiles}, C.},
  \bibinfo{year}{1964}.
\newblock \bibinfo{title}{{The applicability of the third integral of motion:
  Some numerical experiments}}.
\newblock \bibinfo{journal}{\aj} \bibinfo{volume}{69}, \bibinfo{pages}{73}.
%Type = Article
\bibitem[{{Kandrup}(1998)}]{K98}
\bibinfo{author}{{Kandrup}, H.E.}, \bibinfo{year}{1998}.
\newblock \bibinfo{title}{{Phase mixing in time-independent Hamiltonian
  systems}}.
\newblock \bibinfo{journal}{\mnras} \bibinfo{volume}{301},
  \bibinfo{pages}{960--974}.
\newblock \eprint{astro-ph/9809100}.
%Type = Article
\bibitem[{{Kandrup} and {Siopis}(2003)}]{KS03}
\bibinfo{author}{{Kandrup}, H.E.}, \bibinfo{author}{{Siopis}, C.},
  \bibinfo{year}{2003}.
\newblock \bibinfo{title}{{Chaos and chaotic phase mixing in cuspy triaxial
  potentials}}.
\newblock \bibinfo{journal}{\mnras} \bibinfo{volume}{345},
  \bibinfo{pages}{727--742}.
\newblock \eprint{astro-ph/0305198}.
%Type = Article
\bibitem[{{Laskar}(1990)}]{L90}
\bibinfo{author}{{Laskar}, J.}, \bibinfo{year}{1990}.
\newblock \bibinfo{title}{{The chaotic motion of the solar system - A numerical
  estimate of the size of the chaotic zones}}.
\newblock \bibinfo{journal}{\icarus} \bibinfo{volume}{88},
  \bibinfo{pages}{266--291}.
%Type = Article
\bibitem[{{Lega} and {Froeschl{\'e}}(2001)}]{LF01}
\bibinfo{author}{{Lega}, E.}, \bibinfo{author}{{Froeschl{\'e}}, C.},
  \bibinfo{year}{2001}.
\newblock \bibinfo{title}{{On the relationship between fast lyapunov indicator
  and periodic orbits for symplectic mappings}}.
\newblock \bibinfo{journal}{Celest. Mech. Dynam. Astron.} \bibinfo{volume}{81},
  \bibinfo{pages}{129--147}.
%Type = Article
\bibitem[{{Maffione} et~al.(2011){Maffione}, {Darriba}, {Cincotta} and
  {Giordano}}]{MDCG11}
\bibinfo{author}{{Maffione}, N.P.}, \bibinfo{author}{{Darriba}, L.A.},
  \bibinfo{author}{{Cincotta}, P.M.}, \bibinfo{author}{{Giordano}, C.M.},
  \bibinfo{year}{2011}.
\newblock \bibinfo{title}{{A comparison of different indicators of chaos based
  on the deviation vectors: application to symplectic mappings}}.
\newblock \bibinfo{journal}{Celest. Mech. Dynam. Astron.}
  \bibinfo{volume}{111}, \bibinfo{pages}{285--307}.
\newblock \eprint{1108.2196}.
%Type = Article
\bibitem[{{Maffione} et~al.(2013){Maffione}, {Darriba}, {Cincotta} and
  {Giordano}}]{MDCG13}
\bibinfo{author}{{Maffione}, N.P.}, \bibinfo{author}{{Darriba}, L.A.},
  \bibinfo{author}{{Cincotta}, P.M.}, \bibinfo{author}{{Giordano}, C.M.},
  \bibinfo{year}{2013}.
\newblock \bibinfo{title}{{Chaos detection tools: application to a
  self-consistent triaxial model}}.
\newblock \bibinfo{journal}{\mnras} \bibinfo{volume}{429},
  \bibinfo{pages}{2700--2717}.
\newblock \eprint{1212.3175}.
%Type = Article
\bibitem[{{Mahon} et~al.(1995){Mahon}, {Abernathy}, {Bradley} and
  {Kandrup}}]{MABK95}
\bibinfo{author}{{Mahon}, M.E.}, \bibinfo{author}{{Abernathy}, R.A.},
  \bibinfo{author}{{Bradley}, B.O.}, \bibinfo{author}{{Kandrup}, H.E.},
  \bibinfo{year}{1995}.
\newblock \bibinfo{title}{{Transient ensemble dynamics in time-independent
  galactic potentials}}.
\newblock \bibinfo{journal}{\mnras} \bibinfo{volume}{275},
  \bibinfo{pages}{443--453}.
%Type = Article
\bibitem[{{Merritt} and {Fridman}(1996)}]{MF96}
\bibinfo{author}{{Merritt}, D.}, \bibinfo{author}{{Fridman}, T.},
  \bibinfo{year}{1996}.
\newblock \bibinfo{title}{{Triaxial Galaxies with Cusps}}.
\newblock \bibinfo{journal}{\apj} \bibinfo{volume}{460}, \bibinfo{pages}{136}.
\newblock \eprint{arXiv:astro-ph/9511021}.
%Type = Article
\bibitem[{{Muzzio} et~al.(2005){Muzzio}, {Carpintero} and {Wachlin}}]{MCW05}
\bibinfo{author}{{Muzzio}, J.C.}, \bibinfo{author}{{Carpintero}, D.D.},
  \bibinfo{author}{{Wachlin}, F.C.}, \bibinfo{year}{2005}.
\newblock \bibinfo{title}{{Spatial Structure of Regular and Chaotic Orbits in A
  Self-Consistent Triaxial Stellar System}}.
\newblock \bibinfo{journal}{Celest. Mech. Dynam. Astron.} \bibinfo{volume}{91},
  \bibinfo{pages}{173--190}.
%Type = Article
\bibitem[{{Muzzio} and {Mosquera}(2004)}]{MM04}
\bibinfo{author}{{Muzzio}, J.C.}, \bibinfo{author}{{Mosquera}, M.E.},
  \bibinfo{year}{2004}.
\newblock \bibinfo{title}{{Spatial Structure of Regular and Chaotic Orbits in
  Self-Consistent Models of Galactic Satellites}}.
\newblock \bibinfo{journal}{Celestial Mechanics and Dynamical Astronomy}
  \bibinfo{volume}{88}, \bibinfo{pages}{379--396}.
%Type = Article
\bibitem[{{Papaphilippou} and {Laskar}(1998)}]{PL98}
\bibinfo{author}{{Papaphilippou}, Y.}, \bibinfo{author}{{Laskar}, J.},
  \bibinfo{year}{1998}.
\newblock \bibinfo{title}{{Global dynamics of triaxial galactic models through
  frequency map analysis}}.
\newblock \bibinfo{journal}{\aap} \bibinfo{volume}{329},
  \bibinfo{pages}{451--481}.
%Type = Book
\bibitem[{{Press} et~al.(1992){Press}, {Teukolsky}, {Vetterling} and
  {Flannery}}]{PTVF92}
\bibinfo{author}{{Press}, W.H.}, \bibinfo{author}{{Teukolsky}, S.A.},
  \bibinfo{author}{{Vetterling}, W.T.}, \bibinfo{author}{{Flannery}, B.P.},
  \bibinfo{year}{1992}.
\newblock \bibinfo{title}{Numerical Recipes in Fortran 77: The Art of
  Scientific Computing}.
\newblock \bibinfo{publisher}{Cambridge University Press}.
  \bibinfo{edition}{2nd} edition.
%Type = Article
\bibitem[{{S{\'a}ndor} et~al.(2000){S{\'a}ndor}, {{\'E}rdi} and
  {Efthymiopoulos}}]{SEE00}
\bibinfo{author}{{S{\'a}ndor}, Z.}, \bibinfo{author}{{{\'E}rdi}, B.},
  \bibinfo{author}{{Efthymiopoulos}, C.}, \bibinfo{year}{2000}.
\newblock \bibinfo{title}{{The Phase Space Structure Around L4 in the
  Restricted Three-Body Problem}}.
\newblock \bibinfo{journal}{Celest. Mech. Dynam. Astron.} \bibinfo{volume}{78},
  \bibinfo{pages}{113--123}.
%Type = Article
\bibitem[{{S{\'a}ndor} et~al.(2004){S{\'a}ndor}, {{\'E}rdi}, {Sz{\'e}ll} and
  {Funk}}]{SESF04}
\bibinfo{author}{{S{\'a}ndor}, Z.}, \bibinfo{author}{{{\'E}rdi}, B.},
  \bibinfo{author}{{Sz{\'e}ll}, A.}, \bibinfo{author}{{Funk}, B.},
  \bibinfo{year}{2004}.
\newblock \bibinfo{title}{{The Relative Lyapunov Indicator: An Efficient Method
  of Chaos Detection}}.
\newblock \bibinfo{journal}{Celest. Mech. Dynam. Astron.} \bibinfo{volume}{90},
  \bibinfo{pages}{127--138}.
%Type = Article
\bibitem[{{Schwarzschild}(1979)}]{S79}
\bibinfo{author}{{Schwarzschild}, M.}, \bibinfo{year}{1979}.
\newblock \bibinfo{title}{{A numerical model for a triaxial stellar system in
  dynamical equilibrium}}.
\newblock \bibinfo{journal}{\apj} \bibinfo{volume}{232},
  \bibinfo{pages}{236--247}.
%Type = Article
\bibitem[{{Schwarzschild}(1982)}]{S82}
\bibinfo{author}{{Schwarzschild}, M.}, \bibinfo{year}{1982}.
\newblock \bibinfo{title}{{Triaxial equilibrium models for elliptical galaxies
  with slow figure rotation}}.
\newblock \bibinfo{journal}{\apj} \bibinfo{volume}{263},
  \bibinfo{pages}{599--610}.
%Type = Article
\bibitem[{{Skokos}(2001)}]{S01}
\bibinfo{author}{{Skokos}, C.}, \bibinfo{year}{2001}.
\newblock \bibinfo{title}{Alignment indices: a new, simple method for
  determining the ordered or chaotic nature of orbits}.
\newblock \bibinfo{journal}{Journal of Physics A: Mathematical and General}
  \bibinfo{volume}{34}, \bibinfo{pages}{10029--10043}.
%Type = Article
\bibitem[{{Skokos} et~al.(2008){Skokos}, {Bountis} and {Antonopoulos}}]{SBA08}
\bibinfo{author}{{Skokos}, C.}, \bibinfo{author}{{Bountis}, T.},
  \bibinfo{author}{{Antonopoulos}, C.}, \bibinfo{year}{2008}.
\newblock \bibinfo{title}{{Detecting chaos, determining the dimensions of tori
  and predicting slow diffusion in Fermi-Pasta-Ulam lattices by the Generalized
  Alignment Index method}}.
\newblock \bibinfo{journal}{European Physical Journal Special Topics}
  \bibinfo{volume}{165}, \bibinfo{pages}{5--14}.
\newblock \eprint{0802.1646}.
%Type = Article
\bibitem[{{Skokos} et~al.(2007){Skokos}, {Bountis} and {Antonopoulos}}]{SBA07}
\bibinfo{author}{{Skokos}, C.}, \bibinfo{author}{{Bountis}, T.C.},
  \bibinfo{author}{{Antonopoulos}, C.}, \bibinfo{year}{2007}.
\newblock \bibinfo{title}{{Geometrical properties of local dynamics in
  Hamiltonian systems: The Generalized Alignment Index (GALI) method}}.
\newblock \bibinfo{journal}{Physica D Nonlinear Phenomena}
  \bibinfo{volume}{231}, \bibinfo{pages}{30--54}.
\newblock \eprint{0704.3155}.
%Type = Article
\bibitem[{{{\v S}idlichovsk{\'y}} and {Nesvorn{\'y}}(1996)}]{SN96}
\bibinfo{author}{{{\v S}idlichovsk{\'y}}, M.}, \bibinfo{author}{{Nesvorn{\'y}},
  D.}, \bibinfo{year}{1996}.
\newblock \bibinfo{title}{{Frequency modified Fourier transform and its
  applications to asteroids}}.
\newblock \bibinfo{journal}{Celest. Mech. Dynam. Astron.} \bibinfo{volume}{65},
  \bibinfo{pages}{137--148}.
%Type = Article
\bibitem[{{Valluri} and {Merritt}(1998)}]{VM98}
\bibinfo{author}{{Valluri}, M.}, \bibinfo{author}{{Merritt}, D.},
  \bibinfo{year}{1998}.
\newblock \bibinfo{title}{{Regular and Chaotic Dynamics of Triaxial Stellar
  Systems}}.
\newblock \bibinfo{journal}{\apj} \bibinfo{volume}{506},
  \bibinfo{pages}{686--711}.
\newblock \eprint{arXiv:astro-ph/9801041}.
%Type = Article
\bibitem[{{van den Bosch} et~al.(2008){van den Bosch}, {van de Ven}, {Verolme},
  {Cappellari} and {de Zeeuw}}]{VVVCD08}
\bibinfo{author}{{van den Bosch}, R.C.E.}, \bibinfo{author}{{van de Ven}, G.},
  \bibinfo{author}{{Verolme}, E.K.}, \bibinfo{author}{{Cappellari}, M.},
  \bibinfo{author}{{de Zeeuw}, P.T.}, \bibinfo{year}{2008}.
\newblock \bibinfo{title}{{Triaxial orbit based galaxy models with an
  application to the (apparent) decoupled core galaxy NGC 4365}}.
\newblock \bibinfo{journal}{\mnras} \bibinfo{volume}{385},
  \bibinfo{pages}{647--666}.
\newblock \eprint{0712.0113}.
%Type = Article
\bibitem[{{Vogelsberger} et~al.(2008){Vogelsberger}, {White}, {Helmi} and
  {Springel}}]{VWHS08}
\bibinfo{author}{{Vogelsberger}, M.}, \bibinfo{author}{{White}, S.D.M.},
  \bibinfo{author}{{Helmi}, A.}, \bibinfo{author}{{Springel}, V.},
  \bibinfo{year}{2008}.
\newblock \bibinfo{title}{{The fine-grained phase-space structure of cold dark
  matter haloes}}.
\newblock \bibinfo{journal}{\mnras} \bibinfo{volume}{385},
  \bibinfo{pages}{236--254}.
\newblock \eprint{0711.1105}.
%Type = Article
\bibitem[{{Voglis} et~al.(1999){Voglis}, {Contopoulos} and
  {Efthymiopoulos}}]{VCE99}
\bibinfo{author}{{Voglis}, N.}, \bibinfo{author}{{Contopoulos}, G.},
  \bibinfo{author}{{Efthymiopoulos}, C.}, \bibinfo{year}{1999}.
\newblock \bibinfo{title}{{Detection of Ordered and Chaotic Motion Using the
  Dynamical Spectra}}.
\newblock \bibinfo{journal}{Celest. Mech. Dynam. Astron.} \bibinfo{volume}{73},
  \bibinfo{pages}{211--220}.
%Type = Article
\bibitem[{{Voglis} and {Contopoulos}(1994)}]{VC94}
\bibinfo{author}{{Voglis}, N.}, \bibinfo{author}{{Contopoulos}, G.J.},
  \bibinfo{year}{1994}.
\newblock \bibinfo{title}{Invariant spectra of orbits in dynamical systems}.
\newblock \bibinfo{journal}{Journal of Physics A: Mathematical and General}
  \bibinfo{volume}{27}, \bibinfo{pages}{4899--4909}.
%Type = Article
\bibitem[{{Voglis} et~al.(2002){Voglis}, {Kalapotharakos} and
  {Stavropoulos}}]{VKS02}
\bibinfo{author}{{Voglis}, N.}, \bibinfo{author}{{Kalapotharakos}, C.},
  \bibinfo{author}{{Stavropoulos}, I.}, \bibinfo{year}{2002}.
\newblock \bibinfo{title}{{Mass components in ordered and in chaotic motion in
  galactic N-body models}}.
\newblock \bibinfo{journal}{\mnras} \bibinfo{volume}{337},
  \bibinfo{pages}{619--630}.
%Type = Article
\bibitem[{{Zorzi} and {Muzzio}(2012)}]{ZM12}
\bibinfo{author}{{Zorzi}, A.F.}, \bibinfo{author}{{Muzzio}, J.C.},
  \bibinfo{year}{2012}.
\newblock \bibinfo{title}{{Models of cuspy triaxial stellar systems - I.
  Stability and chaoticity}}.
\newblock \bibinfo{journal}{\mnras} \bibinfo{volume}{423},
  \bibinfo{pages}{1955--1963}.
\newblock \eprint{1204.5428}.

\end{thebibliography}

%% not want to use model2-names.bst.



%% \bibitem must have one of the following forms:
%%   \bibitem[Jones et al.(1990)]{key}...
%%   \bibitem[Jones et al.(1990)Jones, Baker, and Williams]{key}...
%%   \bibitem[Jones et al., 1990]{key}...
%%   \bibitem[\protect\citeauthoryear{Jones, Baker, and Williams}{Jones
%%       et al.}{1990}]{key}...
%%   \bibitem[\protect\citeauthoryear{Jones et al.}{1990}]{key}...
%%   \bibitem[\protect\astroncite{Jones et al.}{1990}]{key}...
%%   \bibitem[\protect\citename{Jones et al., }1990]{key}...
%%   \harvarditem[Jones et al.]{Jones, Baker, and Williams}{1990}{key}...
%%


% \end{thebibliography}

\end{document}